\documentclass[conference]{IEEEtran}
\usepackage[latin9]{inputenc}
\usepackage{amsmath}
\usepackage{amssymb}
\usepackage{graphicx}
\usepackage{esint}

\makeatletter
\IEEEoverridecommandlockouts
\usepackage{cite}
\usepackage{amsfonts}\usepackage{algorithmic}
\usepackage{subfigure}
\usepackage{textcomp}
\usepackage{xcolor}
\def\BibTeX{{\rm B\kern-.05em{\sc i\kern-.025em b}\kern-.08em
    T\kern-.1667em\lower.7ex\hbox{E}\kern-.125emX}}
\makeatother

\begin{document}

\title{Traveling Wave Tube Eigenmode Solver for Hot Slow Wave Structure
Based on Particle-In-Cell Simulations}

\author{\IEEEauthorblockN{Tarek Mealy and Filippo Capolino} \IEEEauthorblockA{\textit{Department of Electrical Engineering and Computer Science,
University of California, Irvine, CA 92697, USA} \\
tmealy@uci.edu and f.capolino@uci.edu}}

\IEEEaftertitletext{\noindent A scheme to characterize the dynamics of the electron beam-electromagnetic
power exchange along a traveling wave tube (TWT) is proposed. The
method is based on defining a state vector at discrete periodic locations
along the TWT and determining the transfer matrix of the unit-cell
of the ``hot'' slow-wave structure (SWS) that takes into account
the interaction between the electromagnetic guided field and the electron
beam via particle-in-cell (PIC) simulations. Once the estimate of
the unit-cell transfer matrix is obtained, we show how to find the
hybrid, beam-electromagnetic, eigenmodes in the hot SWS, i.e., where
the electromagnetic guided field interacts with an electron beam,
by using Floquet theory. In particular, we show how do determine the
complex-valued wavenumbers of the hybrid modes and the eigenvectors
associated to them. The method is applied to find the hot modes with
complex wavenumber that can be supported in a TWT amplifier with a
helix SWS. We show dispersion relations of the modal complex wavenumbers
of the hybrid modes when varying frequency and beam voltage; the results
are in agreement with Pierce theory. The method is also applied to
find the complex-wavenumber modes in a hot SWS of a millimeter wave
TWT amplifier based on a serpentine waveguide. The technique is general
and can be applied to any SWS geometry where electromagnetic modes
interact with an electron beam.}

\maketitle
\vphantom{}

\vphantom{}

\section{Introduction}

Traveling wave tube (TWT) amplifiers are the devices of choice for
several decades for radar and satellite communications applications
when high power is required and also when reliability is important,
like in satellite communications \cite{minenna2019traveling,li2018design}.
TWTs are increasingly important to generate high power at millimeter
wave and terahertz frequencies \cite{booske2008plasma,sengele2009microfabrication,booske2011vacuum,gong2011experimental,armstrong2018compact}
where the current technology based on solid state devices struggles
to generate even low power levels. An important component of the TWT
is the slow wave structure (SWS), that is a guiding structure where
the speed of the electromagnetic (EM) wave is reduced to match the
speed of the beam electrons leading to energy transfer from the electron
beam to the EM wave \cite{benford2007high,gilmour1994principles}.
An important mechanism for the energy transfer is the synchronization
of the phase velocity $v_{ph}$ of the EM wave in the SWS with the
average speed of the electrons $u_{0}$. Furthermore, the EM wave
needs to have a longitudinal electric field component $E_{z}$ to
interact with the electron beam to form electron bunches. Therefore
the electron beam is modulated in terms of electron velocity and electron
density forming a ``space-charge wave'' that is synchronized with
the EM wave. The modes in the ``hot'' SWS are complex hybrid physical
phenomena involving both the space-charge wave and EM field, i.e.,
each mode is made of these two components and may have a complex wavenumber.

The study of the ``cold'' eigenmodes in the SWS, i.e., the EM modes
that exist without considering the interaction with the electron beam,
is important to establish the onset of the synchronization condition
between the electron beam's space-charge wave and the EM wave in the
SWS. Denoting with $v_{ph}$ the phase velocity of the EM mode in
the cold SWS and with $u_{0}$ the average velocity of the electron
beam, the initial synchronization condition is $v_{ph}\approx u_{0}$.
There are various EM solvers in commercial software packages that
can be used to find the dispersion diagram of the EM modes in the
cold SWS. Some of the most famous commercial eigenmode solvers are
provided by finite element method-based packages by Ansys HFSS and
DS SIMULIA (previously known as CST Microwave Studio). Often, eigenmode
solvers work under the approximation of a lossless SWS, i.e., the
modes are found in a closed metallic waveguide with perfect electric
conducting walls.

The interaction of an EM wave with the electron beam results in hybrid
(EM+space-charge wave) modes whose phase velocity is different from
the phase velocity of the cold EM eigenmode, hence the ``hot'' eigenmodes,
i.e., the eigenmodes in \textit{interactive} system, have a dispersion
diagram that is different from the one of the cold EM modes, especially
in the frequency region where $v_{ph}\approx u_{0}$. Although the
study of the EM eigenmodes in a cold SWS is very important, the main
operation of TWTs depends mainly on the eigenmodes of the hot SWS.
Note that the modes of the interactive system have complex valued
wavenumbers accounting for possible gain coming from the electron
beam and losses in the metallic waveguide.

The modeling and design of TWTs are carried out by either theoretical
models or simulations. For about the past seventy years, Pierce\textquoteright s
classical small signal theory has been used for the modeling and design
of TWTs \cite{pierce1947theory,pierce1951waves}. Pierce describes
the dispersion relation for hot SWS as cubic polynomial \cite{pierce1947theory}.
Other studies have been provided in the literature to theoretically
model TWTs as in \cite{vainshtein1956electron,sturrock1958kinematics,tamma2014extension,othman2016theory,jassem2020theory}.
Although the theoretical models are considered as good tools for initial
design of TWT, they are inaccurate and the actual performance of TWTs
is assessed by performing accurate simulations.

Advances in 3D electromagnetic simulation software make it possible
to accurately model and simulate complex electromagnetic structures
accounting for the interaction with an electron beam. Most of the
simulation and design work of TWTs is carried out using particle-in-cell
(PIC) codes. A PIC solver is a self-consistent simulation method for
particle tracking that calculates particles trajectories and EM fields
in the time-domain \cite{dawson1983particle,tskhakaya2007particle}.
Some commercial computational software provides solvers based on PIC
code that allows to accurately simulate driven-source problems of
TWTs taking into account all physical aspects of the problem. Although
PIC solvers are currently the most accurate tools to model TWTs, they
require a lot of simulation time and computer memory size, especially
for TWTs with large lengths, therefore, sometimes running multiple
PIC simulations may not be the most practical way to start the optimization
process. Several methods were proposed in literature as an intermediate
step between simple qualitative theoretical models and time-consuming
accurate PIC simulations. Some of thesemethods are based on merging
data extracted from a 3D cold SWS electromagnetic simulator with those
of a particle solver, assuming the SWS is modeled as 1D transmission
line and the beam is modeled as pencil beam with infinitesimal cross-section
area. These methods are used in the well known codes known as CHRISTINE
\cite{antonsen1997christine,antonsen1998traveling}, TESLA \cite{vlasov2002simulation}and
MUSE \cite{wohlbier2002multifrequency}. Other methods based on merging
results from 3D EM simulations and particle solver simulations are
also proposed in \cite{solntsev2015beam,chernyavskiy2016large,chernyavskiy2017modeling,jabotinski2019calculation,minenna2019recent}.

In this paper, we present a method to model TWTs by finding the equivalent
transfer matrix of a unit-cell based on accurate PIC simulations of
a SWS with a relatively small number of unit-cells, which is not very
time consuming. The advantage of extracting the SWS unit-cell transfer
matrix is not only to infer the characteristics of the hybrid modes
of the EM-beam interactive system (the main goal of this paper) but
also to predict the behavior of longer structures without need to
simulate it using PIC (left to future investigations). To date, no
commercial software provides an eigenmode solver for hot SWSs taking
into account the interaction with the electron beam and losses and
the accurate geometry of the SWS.

The method shown in this paper is based on finding the unit-cell transfer
matrix through the interpretation of data extracted from PIC simulation
for relatively short SWSs. We show that the method is used to approximately
calculates the complex wavenumbers of the hybrid modes supported in
a hot SWSs. The method also provides the contribution of the EM wave
and spatial charge wave to each specific hybrid mode associated to
each of the complex modal wavenumbers. The proposed solver is based
on accurate PIC simulations of finite length hot SWSs and takes account
of the precise SWS geometry, materials' EM properties, electron beam
cross-section area, confinement magnetic field and space charge effect.
The proposed solver is based on monitoring both the EM fields and
electron beam dynamics in each unit cell and then find the best transfer
matrix that describe how the hybrid EM field-electron beam propagates
along the TWT as shown in Fig. \ref{Fig:General_Setup}. Some simplifying
assumptions are made, like linearity of EM and space charge waves
excited in the TWT, and the electron beam is assumed not to lose energy
while it travels along the TWT as discussed in the next section. Notably,
the method provides the dispersion diagram of complex valued wavenumbers
versus frequency of all the hybrid modes supported in the interactive
hot SWS.

\section{Theoretical framework}

\begin{figure}
\begin{centering}
\centering \subfigure[]{\includegraphics[width=0.9\columnwidth]{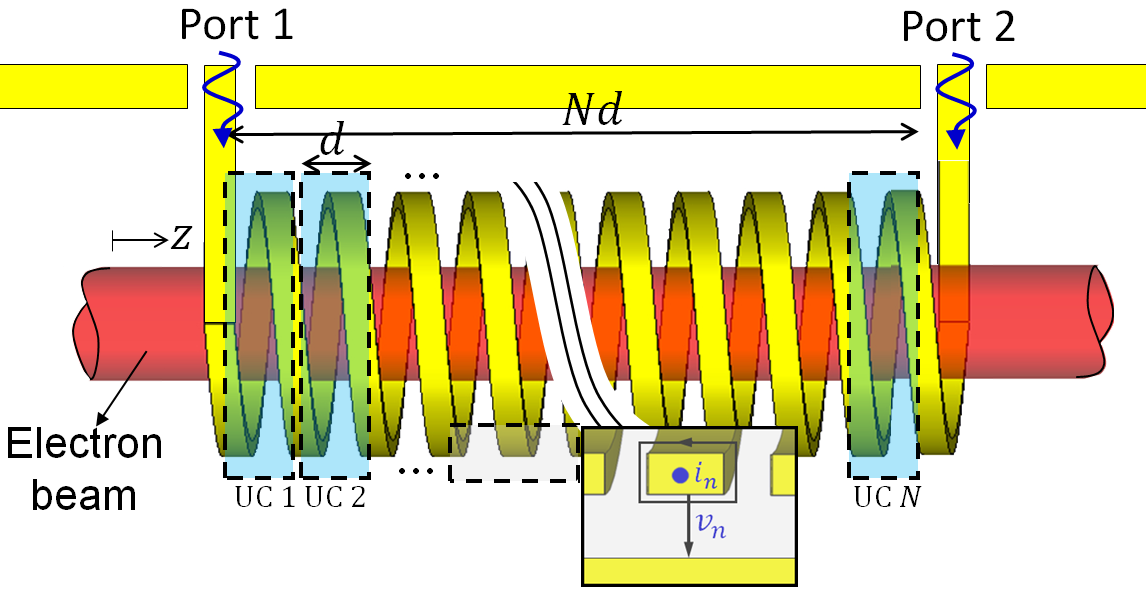}\label{Fig:General_Setup_a}}
\par\end{centering}
\begin{centering}
\centering \subfigure[]{\includegraphics[width=0.9\columnwidth]{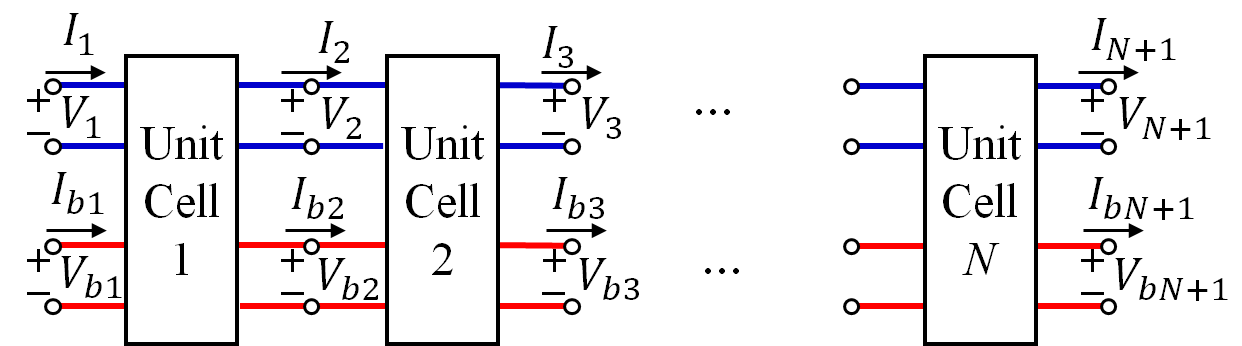}\label{Fig:General_Setup_a_model}}
\par\end{centering}
\begin{centering}
\centering\subfigure[]{\includegraphics[width=0.8\columnwidth]{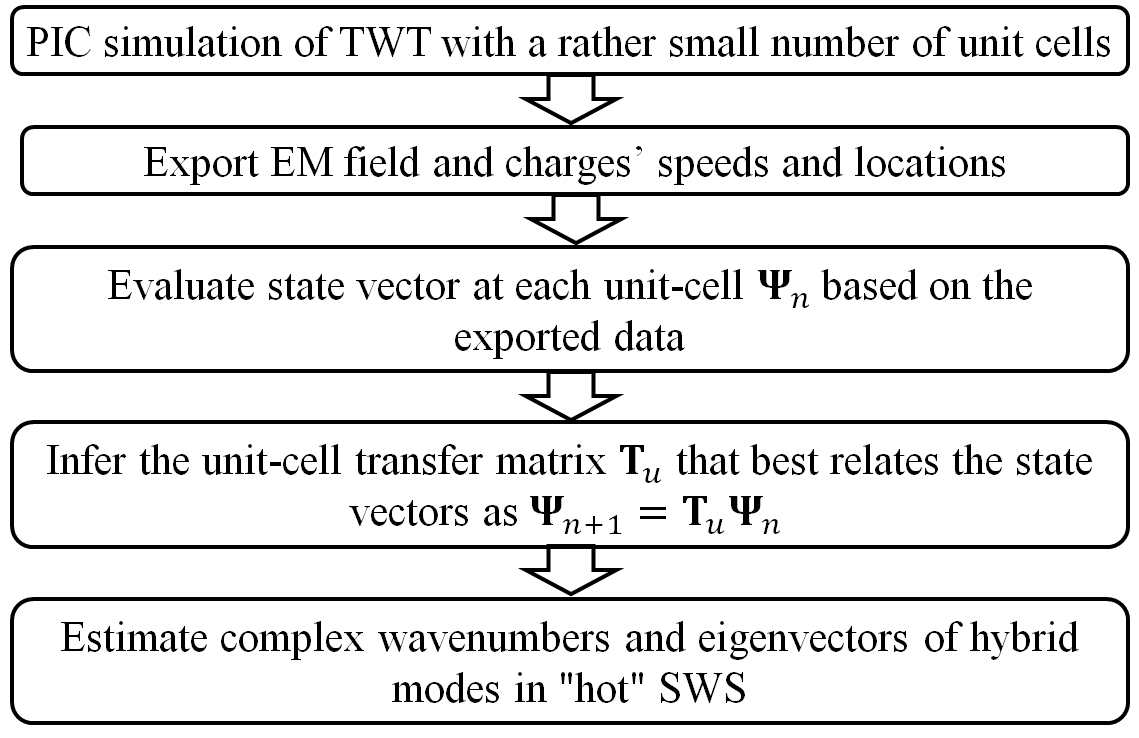}\label{Fig:General_Setup_c}} 
\par\end{centering}
\centering{}\caption{(a) General setup used to determine the complex wavenumber versus
frequency dispersion relation of hybrid modes in hot SWSs based on
PIC simulations. In this figure we show a helix-based SWS, though
the method is general and can be applied to several TWT structures.
In (b) we show the associated circuit model where each unit cell of
the hot SWS is modeled as a multi-port network circuit with \textit{equivalent}
voltages and currents representing EM waves ($V_{n}$, $I_{n}$) and
space-charge waves ($V_{bn}$, $I_{bn}$). In this figure, each unit
cell has 4 ports, two for EM waves (with blue color) and two for space-charge
waves (with red color). The method in this paper calculates the hybrid
modes of the periodic SWS circuit made of periodically cascaded 4-port
network unit cells. (c) Data flowchart used to extract the transfer
matrix $\mathbf{T}_{\mathit{u}}$ of a unit cell and then determine
the hybrid eigenmodes of the hot SWS using the model in (b).\label{Fig:General_Setup}}
\end{figure}

We demonstrate the performance of the proposed eigenmode to be periodically
repeated solver by considering, as an illustrative example, the TWT
made of a helix SWS with period $d$ consisting of $N$ unit-cells
shown in Fig. \ref{Fig:General_Setup_a}. The input and output radio
frequency (RF) signals of the structure are defined as Port 1 and
Port 2. However, in the following both will be used as inputs in order
to find the complex wavenumbers of the hybrid modes. It is important
to point out that the following technique is general and can be applied
to any kind of SWS, hence not only to helix-based SWSs. The PIC solver
simulates the complex interaction between the guided EM wave and electron
beam using a large number of charged particles and it follows their
trajectories in self-consistent electromagnetic fields computed on
a fixed mesh.

\subsection{Assumptions\label{subsec:Assumptions}}

The goal is to estimate the wavenumbers and composition of the hybrid
modes of the interactive system made of EM field and the electron
beam in a TWT amplifier as the one in Fig. \ref{Fig:General_Setup_a}.
The method is based on finding the modes supported by the periodic
distribution of equivalent networks shown in Fig. \ref{Fig:General_Setup_a_model}.
Each unit cell is modeled using a transfer matrix which is calculated,
as explained in this section, based on the results of accurate time
domain PIC numerical simulations of the guided EM field interacting
with the electron beam. The PIC method isbased oncharged particles
motion equations and EM fields, discretized in space and time. Therefore,
the EM field satisfies a discretized form of the time-domain Maxwell
equations, and the PIC simulator accounts for the precise geometry
and materials of the SWS and for the EM boundary conditions on the
lateral walls of the periodic waveguide. Once time domain data are
extracted from a PIC simulator, hybrid modes are found by imposing
periodic boundary conditions along the TWT longitudinal direction
\textit{z} in the phasor domain using the periodic distribution of
equivalent unit cell networks in Fig. \ref{Fig:General_Setup_a_model},
as discussed later on in this section. Therefore, the method estimates
the complex wavenumbers of the hybrid EM-beam modes that exist in
the infinitely-long sequence of equivalent unit cell networks in Fig.
\ref{Fig:General_Setup_a_model} by elaborating the data provided by
the PIC method, simulating the EM field interacting with electrons
in a realistic TWT of finite length. In Fig. \ref{Fig:General_Setup_c}
we show the data flowchart used to accomplish this task: the time
domain data extracted from PIC simulations of relatively short SWSs
are transformed into phasors and then used to find the unit-cell transfer
matrix $\mathbf{T}_{\mathit{u}}$ of which we find the eigenvalues
and eigenvectors as described in this section. 

In using and elaborating the data provided by the PIC solver, we make
the following assumptions. Although a PIC solver calculates the speeds
of discrete charged particles, we represent the longitudinal speed
of all electron-beam charges as a one dimensional (1D) function $u_{b}^{tot}(z,t)$.
This is achieved by averaging the speed of the charges at each \textit{z}-cross
section as described later on. Therefore, the electron-beam velocity
and density are described by the functions $u_{b}^{tot}(z,t)=u_{0}+u_{b}(z,t)$,
and $\rho_{b}^{tot}(z,t)=\rho_{0}+\rho_{b}(z,t)$, where $u_{0}$
and $\rho_{0}$ are the velocity and the density of the electrons
in the unperturbed beam (i.e., the dc parts), and $u_{b}(z,t)$ and
$\rho_{b}(z,t)$ represent their modulation functions. In what follows,
the structure is excited by monochromatic EM signals, hence we assume
that the beam modulations $u_{b}(z,t)$ and $\rho_{b}(z,t)$ are also
monochromatic. We also assume that the ac modulation of the electron
beam is small compared to the dc part, therefore the electron beam
current is well approximated by the function $i_{b}^{tot}(z,t)=-I_{0}+i_{b}(z,t)$,
where $I_{0}$ is the dc value and $i_{b}(z,t)$ is its time harmonic
modulation, hence we neglect non linear effects. These assumptions
are the same as in the Pierce model \cite{pierce1947theory,pierce1951waves},
but the ac values are here calculated using averaging of results taken
from time-domain PIC simulations as described later on in this section.
All the calculations are based on the steady state regime in a TWT,
therefore the time domain signals calculated by PIC are transformed
into phasors thanks to the assumption that every ac quantity is sinusoidal.

We assume that the EM fields in the \textit{hot} SWS of finite length
are represented in phasor domain as superposition of modes of the
infinitely-long hot SWS as

\begin{equation}
\begin{array}{c}
\mathbf{E}(x,y,z)=\sum_{m}\mathbf{E}^{\mathrm{mode}m}(x,y,z),\\
\mathbf{H}(x,y,z)=\sum_{m}\mathbf{\mathbf{H}^{\mathrm{mode}\mathit{m}}}(x,y,z),
\end{array}\label{eq:EMModeSuperpos}
\end{equation}
where $\mathbf{E}^{\mathrm{mode}m}$ and $\mathbf{H}^{\mathrm{mode}m}$
are the electric and magnetic fields of the $m^{th}$ hybrid mode
in the infinitely-long hot structure and they are assumed to be represented
as a summation of Floquet spatial harmonics as

\begin{equation}
\begin{array}{c}
\mathbf{E}^{\mathrm{mode}m}=e^{-jk^{\mathrm{mode}m}z}\sum_{q}\mathbf{e}^{\mathrm{mode}m}(x,y)e^{-j2\pi qz/d},\\
\mathbf{H}^{\mathrm{mode}m}=e^{-jk^{\mathrm{mode}m}z}\sum_{q}\mathbf{h}^{\mathrm{mode}m}(x,y)e^{-j2\pi qz/d},
\end{array}\label{eq:FloquetModes}
\end{equation}
indeed they are sampled with a spatial period \textit{d} along the
SWS. Our goal is to find the complex-valued wavenumbers $k^{\mathrm{mode}m}$
and the eigenvectors of the hybrid beam-electromagnetic eigenmodes
of the infinitely-long \textquotedblleft hot\textquotedblright{} structure
using PIC simulations of the finite-length structure. The discussion
in the rest of the paper is based on linearity of the system with
respect to the ac EM and space charge waves, and on the assumption
that the electron beam does not lose energy along its travel along
the TWT, therefore the dc electron beam velocity $u_{0}$ is kept
constant along the TWT. This assumption is important because we assume
that the transfer matrix describing each periodic cell is the same
along the whole SWS length.

The physical quantities that represent the EM modes in the interacting
SWS are electric and magnetic fields $\mathbf{E}(x,y,z,t)$ and $\mathbf{H}(x,y,z,t)$
which are represented in terms of equivalent voltages and currents
$v_{n}(t)$ and $i_{n}(t)$ at discrete location of the periodic structure,
where $n$ is the unit cell index number. In the rest of the paper
we assume that only one cold EM mode is able to propagate in each
direction of the cold SWS, hence only a single pair $(v_{n},$$i_{n})$
will be sufficient to describe the EM wave. Although these quantities
cannot be uniquely defined in most kinds of waveguides, it is possible
to define them and use them to model the space and temporal dynamics
in a waveguide \cite{marcuvitz1951waveguide,felsen1994radiation,collin1990field}.
The space charge wave is represented by equivalent ac kinetic voltages
and beam currents $v_{bn}(t)$ and $i_{bn}(t)$, respectively, using
the averaging method in Fig. \ref{Fig:Particles_Set_Show} as described
next. We define a state vector that describes the EM and space-charge
waves at locations $z=z_{n}=nd$ as

\begin{equation}
\mathbf{\boldsymbol{\psi}}_{n}(t)=[\begin{array}{cccc}
v_{n}(t), & i_{n}(t), & v_{bn}(t), & i_{bn}(t)\end{array}]^{T}.\label{eq:TD-StateVect}
\end{equation}

In the following sections we explain how this state vector is calculated
using PIC calculations at $z=nd$ locations along the SWS. Since we
assume that at steady state all the quantities involved in the state
vector are monochromatic, we define a state vector in phasor-domain
as

\begin{equation}
\mathbf{\Psi}_{n}=[\begin{array}{cccc}
V_{n}, & I_{n}, & V_{bn}, & I_{bn}\end{array}]^{T},\label{eq:FD-StateVect}
\end{equation}
assuming an implicit $e^{j\omega t}$ time dependence. 

In the phasor domain we assume that the longitudinal propagation of
the state vector satisfies the equation $\mathbf{\Psi}_{n+1}=\mathbf{T}_{\mathit{u}}\mathbf{\Psi}_{n}$,
where $\mathbf{T}_{\mathit{u}}$ is the periodic unit cell transfer
matrix, which is unknown and assumed invariant along the periodic
cells of the SWS. As explained next, the first goal is to provide
a method to estimate the transfer matrix $\mathbf{T}_{\mathit{u}}$.
Then, we assume that each of the hybrid EM-charge wave mode in (\ref{eq:EMModeSuperpos})-(\ref{eq:FloquetModes})
is described by a state vector variation as $\mathbf{\Psi}_{n}\mathbf{=\Psi}^{\mathrm{mode}m}e^{-jk_{m}nd}$,
where $k_{m}$ is the complex wavenumber of such a mode. The goal
of this paper is to find the complex wavenumbers $k_{m}$ of all the
hybrid modes in the hot SWS and to find the EM and beam modal weights
in (\ref{eq:FD-StateVect}) for each of the hybrid modes. The goal
is achieved by solving the eigenvalue problem

\begin{equation}
\mathbf{T}_{\mathit{u}}\mathbf{\Psi}^{\mathrm{mode}m}=e^{-jk_{m}d}\mathbf{\Psi}^{\mathrm{mode}m},\label{eq:EigenProb}
\end{equation}
once the estimate of the transfer matrix $\mathbf{T}_{\mathit{u}}$
has been calculated as described in the next two sections.

\subsection{Determination of the system state-vector}

\begin{figure}
\centering{}\includegraphics[width=0.65\columnwidth]{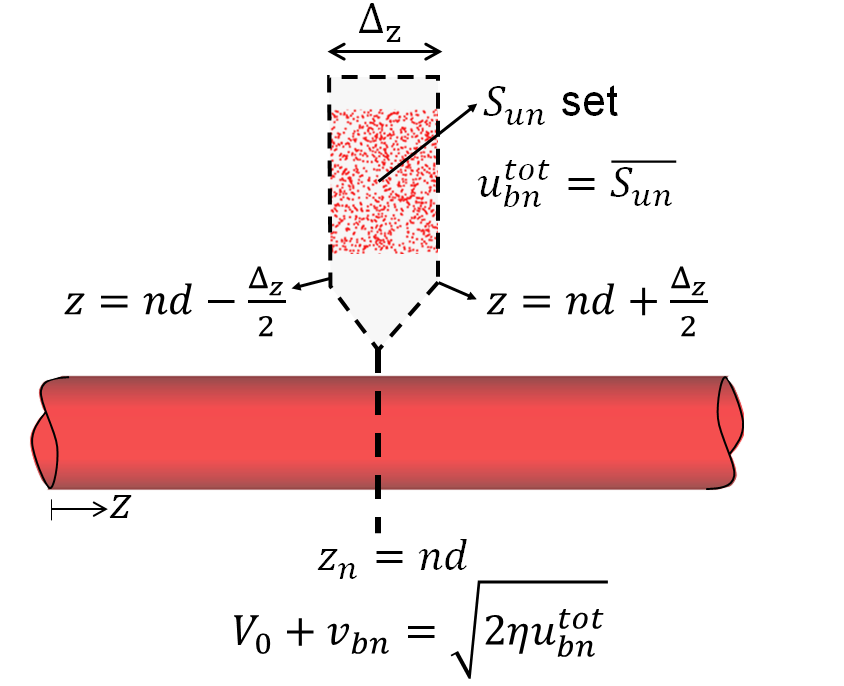}\caption{Illustration of how the speed of the space-charge wave $u_{bn}^{tot}$
is calculated at the entrance of the $n^{th}$ unit-cell (at $z=nd$)
using PIC simulation data. The space-charge wave speed at $z=nd$
is calculated as the average of the speeds of the PIC-defined charges
$S_{u}(t)$ that are in the proximity of $z=z_{n}=nd$, i.e., in the
small range defined as $z_{n}-\Delta_{z}/2<z_{c}(t)<z_{n}+\Delta_{z}/2$,
at time \textit{t}. The subset of all the PIC-defined charges in this
spatial interval at time \textit{t} is called $S_{un}(t)$ because
are about to cross the section at $z=z_{n}$. Since the set of $S_{un}(t)$
is composed of many PIC-defined charges, we define their collection
average $u_{bn}^{tot}=\overline{S_{un}}$, and the charge-wave equivalent
kinetic voltage is $v_{bn}^{tot}=\sqrt{2\eta u_{bn}^{tot}}.$\label{Fig:Particles_Set_Show}}
\end{figure}

For the particular illustrative example shown in Fig. \ref{Fig:General_Setup_a},
we define the voltage as the electric potential difference between
helix loops and the host waveguide, as a function of the electric
field via $v_{n}(t)=\int\boldsymbol{\mathbf{e}}(t)\cdot\mathbf{d}\mathrm{\boldsymbol{\ell}}_{n}$,
where the index $n$ here represents the $n^{th}$ unit-cell and $\mathbf{d}\boldsymbol{\ell}_{n}$
is the incremental vector length along the path between the $n^{th}$
helix loop and the host waveguide in the $n^{th}$ unit-cell as shown
in Fig. \ref{Fig:TWT_Helix_Geom_a}. An alternative way to define
the voltage for the helix is presented in Appendix A. The current
that equivalently represents the EM mode can be defined as the physical
current flowing in the helix tape wire which is determined from the
magnetic field using the integral $i_{n}(t)=\oint\mathbf{h}(t)\cdot\mathbf{d}\mathbf{C}_{n}$,
along the path $C_{n}$ around a helix wire of the $n^{th}$ unit-cell
as shown in Fig. \ref{Fig:General_Setup_a} (using the projection
of this current along the \textit{z} direction leads to the same result).
It is important to point out that equivalent voltage and current representing
the electric and magnetic fields can be similarly defined also for
other SWSs (such as serpentine waveguides, overmoded waveguides, etc.)
using the equivalent field representation described in \cite{collin1990field,felsen1994radiation}.

The PIC solver provides the speeds and locations of all the charged
particles used to model the electron beam at any time $t$. We define
two dynamic sets that involve the speeds and coordinates $z$ of all
the charges in PIC simulation at anytime instant $t$ as $S_{u}(t)=\left\{ u_{1}(t),u_{2}(t),...,u_{Nq}(t)\right\} $
and $S_{z}(t)=\left\{ z_{1}(t),z_{2}(t),...,z_{N_{q}(t)}(t)\right\} $,
respectively, where $N_{q}(t)$ is the total number of charged particles
used by the PIC simulator to model the electron beam at time instant
$t$. The space-charge wave modulating the electron beam is assumed
to be represented using two physical quantities: the electrons speed
which is expressed in term of the beam equivalent kinetic voltage
$v_{b}^{tot}(z,t)$, and the space-charge wave current modulation
$i_{b}^{tot}(z,t)$, as also described in Refs. \cite{tsimring2006electron,gilmour2011klystrons,pierce1951waves}.
The beam total equivalent kinetic voltage at the entrance of the $n^{th}$
unit-cell is defined as $v_{bn}^{tot}(t)=\sqrt{2\eta u_{bn}^{tot}(t)},$
where $u_{bn}^{tot}(t)=\overline{S_{un}(t)}$ is the equivalent speed
of the space-charge wave calculated as the average of the speeds of
the set 

\begin{equation}
S_{un}(t)=\left\{ \begin{array}{ccc}
S_{u}(t) & | & nd-\dfrac{\Delta_{z}}{2}<S_{z}(t)<nd+\dfrac{\Delta_{z}}{2}\end{array}\right\} ,
\end{equation}
that is defined by all the charged particles that are in vicinity
of the entry of the $n^{th}$ unit-cell ($z=nd$), and within the
small spatial interval $\Delta_{z}$, as illustrated in Fig. \ref{Fig:Particles_Set_Show}.
The length of the spatial interval $\Delta_{z}$ is chosen to be very
small, i.e, $\Delta_{z}\leq\lambda_{0b}/20$, where $\lambda_{0b}=u_{0}/f$,
and $u_{0}$ is the electron time-average speed and $f$ is the frequency
modulating the space-charge wave. Although $\Delta_{z}$ is chosen
to be small, it should be also large enough to contain a very large
set of charged particles, as illustrated in Fig. \ref{Fig:Particles_Set_Show}.
The charge-wave current at the $n^{th}$ unit-cell ($z=nd$) is defined
as $i_{bn}^{tot}(t)=-\rho_{bn}^{tot}(t)u_{bn}^{tot}(t)$, where $\rho_{bn}^{tot}(t)$
is the electron beam charge density at the entry of the $n^{th}$
unit-cell and is calculated as $\rho_{bn}^{tot}(t)=q_{e}N_{bn}(t)/\Delta_{z}$
where $q_{e}$ is the charge value of each PIC-defined particle and
$N_{bn}(t)$ is the number of such charged particles in the set $S_{un}(t)$.
The ac modulated parts are then calculated as $v_{bn}(t)=v_{bn}^{tot}(t)-V_{0}$
and $i_{bn}(t)=i_{bn}^{tot}(t)+I_{0}$ which are used later on to
construct the system's state vector. As described below, when we calculate
the phasor $I_{bn}$ associated to $i_{bn}(t)$ we will retain only
the frequency component at radian frequency $\mathbf{\boldsymbol{\omega}}$
and not the higher order harmonics.

It is important to point out that in this illustrative example we
assume that the SWS supports one cold EM mode that can propagate in
each direction and that the electron beam is represented by a single
state that describes the average behavior of the speed and density
of the charged particles distribution. Describing the EM-chanrge wave
state using the four-dimensional state vector (\ref{eq:TD-StateVect})constitutes
a good approximation in many cases where the SWS supports only one
EM mode (in each direction) and the electron beam is modulated in
a homogeneous way, i.e., the beam modulation does not change with
radial and azimuthal angular directions. However, a more accurate
model of the hot SWS could be obtained by using an equivalent multi-transmission
line model to describes all the EM modes in the SWS, and an equivalent
multi ``beam transmission line'' (or multi stream beam) to describe
the electron beam. Indeed, since we know that in reality the momentum
and charge density description of the electron beam dynamics usually
looks like a multi-valued function, it may be convenient to decompose
the electron beam using various areas in transverse cross section
leading to a multi ``beam transmission line'' with multiple kinetic
voltages and space-charge wave currents. For the sake of simplicity,
in this paper the electron beam dynamics is represented only with
one ``beam transmission line'', i.e., with a single ($v_{b}(t),i_{b}(t)$)
pair.

The interaction between the space-charge wave and the EM wave yields
three eigenmodes that travel in the beam direction in addition to
a mode (mainly made of only EM field) propagating opposite to the
beam direction, indeed the latter has very little interaction with
the electron beam \cite{pierce1951waves}. The three hybrid modes
with positive phase velocity are composed of both EM fields and space-charge
wave modulations, and form the ``three-wave'' model used in Refs.
\cite{pierce1947theoryTWT,pierce1951waves}. Under the assumption
of using a single tone excitation of an EM wave from Port 1 and/or
Port 2, all four hybrid EM-charge wave modes in the interacting system
can be excited: an excitation from Port 1 mainly excites the three
interacting hybrid modes, whereas the excitation from Port 2 excites
mainly the EM mode propagating in opposite direction of the beam.
Reflections may occur at the left and right ends in a realistic finite-length
SWS, so in reality all four modes may be present, depending on the
EM reflection coefficients at the two ends. 

At steady state, the state vector is represented in phasor-domain
as in (\ref{eq:FD-StateVect}), assuming an implicit $e^{j\omega t}$
time dependence for all physical quantities. The phases of phasors
are calculated with respect to a fixed time at steady state. The phasor-domain
representation (\ref{eq:FD-StateVect}) of the state vector $\mathbf{\boldsymbol{\psi}}_{n}(t)$
is calculated as

\begin{equation}
\mathbf{\Psi}_{n}=\dfrac{1}{T}\int_{t=0}^{t=T}\mathbf{\boldsymbol{\psi}}_{n}(t+t_{ref})e^{-j\omega t}dt,\label{eq:phasor_Trans}
\end{equation}
where $T=2\pi/\omega$ and $t_{ref}$ is a time reference used to
calculate the phasors and it should be greater than the steady state
time, i.e., $t_{ref}>t_{ss}.$ It is important to mention that only
the fundamental component at frequency $\mathbf{\boldsymbol{\omega}}$
is maintained after the Fourier transform is carried out to build
the state vector in phasor domain $\mathbf{\Psi}_{n}$ in (\ref{eq:FD-StateVect}),
hence all frequency harmonics in $\mathbf{\boldsymbol{\psi}}_{n}(t)$
are neglected in the following. Lower case letters are used for the
time-domain representation whereas capital letters are used for the
phasor-domain representation. In the phasor-domain, we model each
unit cell of the interacting SWS as a 4-port network circuit with
voltages and currents representing both the EM waves and the electron
beam dynamics, as shown in Fig. \ref{Fig:General_Setup_a}. As described
in the previous section, under the assumption of small signal modulation
of the beam\textquoteright s electron velocity and charge density,
the 4-port networks modeling the EM-charge wave interaction in each
unit-cell of the hot SWS are assumed identical. Therefore, one can
define a $4\times4$ transfer matrix $\mathbf{T}_{\mathit{u}}$ of
the interaction unit-cell of a SWS using the relation between the
input and output state vector at each unit-cell as

\begin{equation}
\begin{array}{c}
\mathbf{\Psi}_{2}=\mathbf{T}_{\mathit{u}}\mathbf{\Psi}_{1},\ \ \ \ \ \ (\ref{eq:Tmatrix-StateVector}.1)\\
\mathbf{\Psi}_{3}=\mathbf{T}_{\mathit{u}}\mathbf{\Psi}_{2},\ \ \ \ \ \ (\ref{eq:Tmatrix-StateVector}.2)\\
\vdots\ \ \ \ \ \ \ \ \ \ \ \ \ \ \ \\
\mathbf{\Psi}_{N+1}=\mathbf{T}_{\mathit{u}}\mathbf{\Psi}_{N},\ \ \ \ \ \ (\ref{eq:Tmatrix-StateVector}.N)
\end{array}\label{eq:Tmatrix-StateVector}
\end{equation}
where $\mathbf{\Psi}_{n+1}$ and $\mathbf{\Psi}_{n}$ are the input
and output state vectors of the $n^{th}$ unit-cell, respectively,
where \textit{n} = 1, 2,..\textit{ N}. The state vectors $\mathbf{\Psi}_{n}$
are calculated from $\mathbf{\psi}_{n}$ using PIC simulations; then
an estimate of the transfer matrix $\mathbf{T}_{\mathit{u}}$ is inferred
by the method described in the following section.

\subsection{Finding the transfer matrix of a unit cell of the interactive system}

\subsubsection{Approximate best fit solution}

The relations in (\ref{eq:Tmatrix-StateVector}) represent $4N$ linear
equations in $16$ unknowns which are the elements of the transfer
matrix $\mathbf{T}_{u}$. The system in (\ref{eq:Over_sys}) is mathematically
referred to as overdetermined because the number of linear equations
($4N$ equations) is greater than the number of unknowns (16 unknowns).
We rewrite (\ref{eq:Tmatrix-StateVector}) by clustering all the given
equations in matrix form as 
\begin{equation}
\left[\mathbf{W}_{2}\right]_{4\times N}=\left[\mathbf{T}_{\mathit{u}}\right]_{4\times4}\left[\mathbf{W}_{1}\right]_{4\times N}\label{eq:Over_sys}
\end{equation}
where
\begin{equation}
\mathbf{W}_{1}=\left[\begin{array}{cccc}
\mathbf{\Psi}_{1}, & \mathbf{\Psi}_{2}, & \ldots & \mathbf{\Psi}_{N}\end{array}\right]
\end{equation}
is a $4\times N$ matrix and its columns are the state vectors at
input of each unit-cell, and

\begin{equation}
\mathbf{W}_{2}=\left[\begin{array}{cccc}
\mathbf{\Psi}_{2}, & \mathbf{\Psi}_{3}, & \ldots & \mathbf{\Psi}_{N+1}\end{array}\right]
\end{equation}
is an analogous $4\times N$ matrix but with a shifted set of the
state vectors, i.e., its columns are the state vectors at the output
of each unit-cell. Our first goal is to find the 16 elements of the
transfer matrix $\mathbf{T}_{\mathit{u}}$.

An approximate solution that best satisfies all the given equations
in Eq. (\ref{eq:Tmatrix-StateVector}), i.e., minimizes the sums of
the squared residuals, $\sum_{n}\left|\left|\mathbf{\Psi}_{n+1}-\mathbf{T}_{\mathit{u}}\mathbf{\Psi}_{n}\right|\right|$
is determined similarly as in \cite{forsythe1977computer,williams1990overdetermined,anton2013elementary}
and is given by

\begin{equation}
\mathbf{T}_{\mathit{u,best}}=\left(\left[\mathbf{W}_{2}\right]_{4\times N}\left[\mathbf{W}_{1}\right]_{4\times N}^{T}\right)\left(\left[\mathbf{W}_{1}\right]_{4\times N}\left[\mathbf{W}_{1}\right]_{4\times N}^{T}\right)^{-1}.\label{eq:Best_Fit}
\end{equation}

It is important to point out that all the four modes of the interactive
EM-charge wave system should be excited to be able to have four independent
columns in the construction of the matrices $\mathbf{W}_{1}$ and
$\mathbf{W}_{2}$ since we need apply the inverse operation in (\ref{eq:Best_Fit}).
This occurs when there is sufficient amount of power incident on Port
1 and Port 2.

\subsubsection{Distinct determined solutions}

The transfer matrix $\mathbf{T}_{\mathit{u}}$ can also be determined
directly by taking any four equations of Eq.(\ref{eq:Tmatrix-StateVector}),
assume we choose Eq. (\ref{eq:Tmatrix-StateVector}.$q)$, Eq. (\ref{eq:Tmatrix-StateVector}.$i)$,
Eq. (\ref{eq:Tmatrix-StateVector}.$j)$ and Eq. (\ref{eq:Tmatrix-StateVector}.$k)$,
and therefore yields

\begin{equation}
\mathbf{T}_{\mathit{u,qijk}}=\left[\mathbf{w}_{2,qijk}\right]_{4\times4}\left[\mathbf{w}_{1,qijk}\right]_{4\times4}^{-1},\label{eq:small_sys}
\end{equation}
where

\begin{equation}
\mathbf{w}_{1,qijk}=\left[\begin{array}{cccc}
\mathbf{\Psi}_{q}, & \mathbf{\Psi}_{i}, & \mathbf{\Psi}_{j}, & \mathbf{\Psi}_{k}\end{array}\right]
\end{equation}
and

\begin{equation}
\mathbf{w}_{2,qijk}=\left[\begin{array}{cccc}
\mathbf{\Psi}_{q+1}, & \mathbf{\Psi}_{i+1}, & \mathbf{\Psi}_{j+1}, & \mathbf{\Psi}_{k+1}\end{array}\right].
\end{equation}

Assuming the SWS has $N$ unit cells, there will be $C_{4}^{N}$ possible
solutions for $\mathbf{T}_{u,qijk}$ , where $C_{4}^{N}=N!/((N-4)!4!)$
is the number of the combinations to choose $q,i,j$ and $k$ out
of $N$ choices. Under the assumption that the transfer matrices of
each unit-cell are identical, the sets of four eigenvalues resulting
from the $C_{4}^{N}$ solutions of $\mathbf{T}_{u,qijk}$ should be
identical too. However, the electron beam non-linearity and other
factors may cause small discrepancy in the eigenvalues resulting from
the various eigenmode solutions of $\mathbf{T}_{u,qijk}$ , as shown
in the next section.

It is important to point out that some combinations may result in
a underdetermined system where the rank of $\mathbf{w}_{1,qijk}$
could be less than 4, i.e., $\mathbf{w}_{1,qijk}$ could be singular.
For example, when selecting unit cells toward the right end of the
TWT (e.g., $q=N-3,i=N-2,j=N-1$ and $k=N$), the state vectors forming
$\mathbf{w}_{1.qijk}$ are dominated by only one mode that has exponential
growing in $z$ direction, and therefore the matrix $\mathbf{w}_{1,qijk}$
tends to be singular. Therefore, one can neglect combinations that
are close to be singular by checking the determinant of $\mathbf{w}_{1,qijk}$
for each combination. Following this method, multiple transfer matrices
are found that lead to multiple wavenumbers that are clustered around
four complex values.

\subsection{Finding the hybrid eigenmodes of the interactive system}

Once the transfer matrix is estimated (either using the best-approximate
solution $\mathbf{T}_{\mathit{u,best}}$ of the overdetermined system
or determined solutions $\mathbf{T}_{u,qijk}$), the hybrid eigenmodes
are determined by assuming a state vector has the form of $\mathbf{\Psi_{\mathrm{\mathit{n}}}}\propto e^{-jknd}$,
where $k$ is the complex Bloch wavenumber that has to be determined
and $d$ is the SWS period. Inserting the assumed sate vector \textit{z}-dependency
in (\ref{eq:Tmatrix-StateVector}) yields the eigenvalue problem in
(\ref{eq:EigenProb}). Note that eigenvalues $e^{-jkd}$ and eigenvectors
$\mathbf{\Psi}_{n}$ of the eigenvalue problem in (\ref{eq:EigenProb})
depend only on the transfer matrix $\mathbf{T}_{u}$. The four eigenvalues,

\begin{equation}
e^{-jkd}=\mathrm{eig}(\mathbf{T}_{\mathit{u}}),\label{eq:Eigenvalue}
\end{equation}
of the transfer matrix $\mathbf{T}_{\mathit{u}}$ lead to four Floquet-Bloch
modes $k_{m}$, where $m=1,2,3,4$, with harmonics $k_{m}+2\pi q/d$,
where $q$ is an integer that defines the Floquet-Bloch harmonic index
as in (\ref{eq:FloquetModes}). Some examples are provided in the
next sections. Note that (\ref{eq:EigenProb}) provides also the eigenvectors
$\mathbf{\Psi}^{\mathrm{mode}m}$ and important information can be
extracted from them. Each $m^{th}$ eigenvector possesses the information
of the respective weights of the EM field ($V,I$) and space-charge
wave ($V_{b},I_{b}$) in making that particular hybrid eigenmode solution.
Furthermore, including the case when more EM modes are used in the
SWS interaction zone or when two hybrid modes concur in the synchronization,
an analysis of the eigenvectors can also show possible eigenvector
degeneracy conditions. For example, in \cite{mealy2019backward,mealy2019exceptional,mealy2020exceptional},
two hybrid modes are fully degenerate in wavenumbers and eigenvectors
forming what was called a ``degenerate synchronization'' (degeneracy
between two hot modes). Other important degeneracy conditions are
those studied in \cite{abdelshafy2018electron}, where three or four
fully degenerate EM modes in the cold SWS are used in the synchronization
with the electron beam, a condition refer to as ``multimode synchronization''
(degeneracy among cold EM modes).

In a finite length TWT, the total EM field (represented by $V_{n}^{\mathrm{tot}}$,
$I_{n}^{\mathrm{tot}}$ ) and space-charge wave (represented by $V_{bn}^{\mathrm{tot}}$,
$I_{bn}^{\mathrm{tot}}$) resulting from their interaction, calculated
at each $n^{th}$ location, are represented in terms of the four eigenmodes,

\begin{equation}
\mathbf{\Psi}_{n}^{\mathrm{tot}}=\sum_{m=1}^{4}a_{m}\mathbf{\Psi}^{\mathrm{mode}m}e^{-jk_{m}nd}\label{eq:ModeSuperpos}
\end{equation}
where $a_{m}$ is the weight of the $m^{th}$ mode, which depends
on the mode excitation and boundary conditions, and $\mathbf{\Psi}^{\mathrm{mode}m}$
is the interactive system eigenvectors obtained from (\ref{eq:EigenProb}).
Each Floquet-Bloch mode in the periodic hot SWS is represented as
$\mathbf{\Psi}^{\mathrm{mode}m}e^{-jk_{m}nd}$.

In this paper we show how to determine the eigenvector $\mathbf{\Psi}^{\mathrm{mode}m}$
and wavenumber $k_{m}$ of each of the four hybrid eigenmodes ($m=1,2,3,4$),
using two illustrative examples: a centimeter wave (i.e., ``microwave'')
and a millimeter wave TWT amplifier.

\section{Application to a helix-based TWT amplifier\label{sec:Helix_TWT}}

\begin{figure}
\begin{centering}
\subfigure[]{\includegraphics[width=0.9\columnwidth]{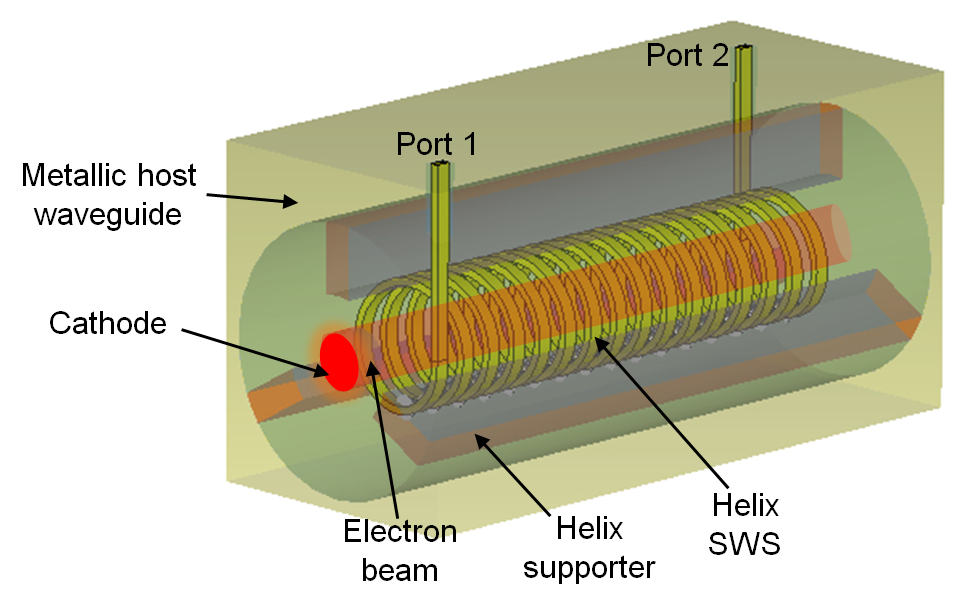}\label{Fig:TWT_Helix_Geom_a}}
\par\end{centering}
\centering{}\subfigure[]{\includegraphics[width=0.8\columnwidth]{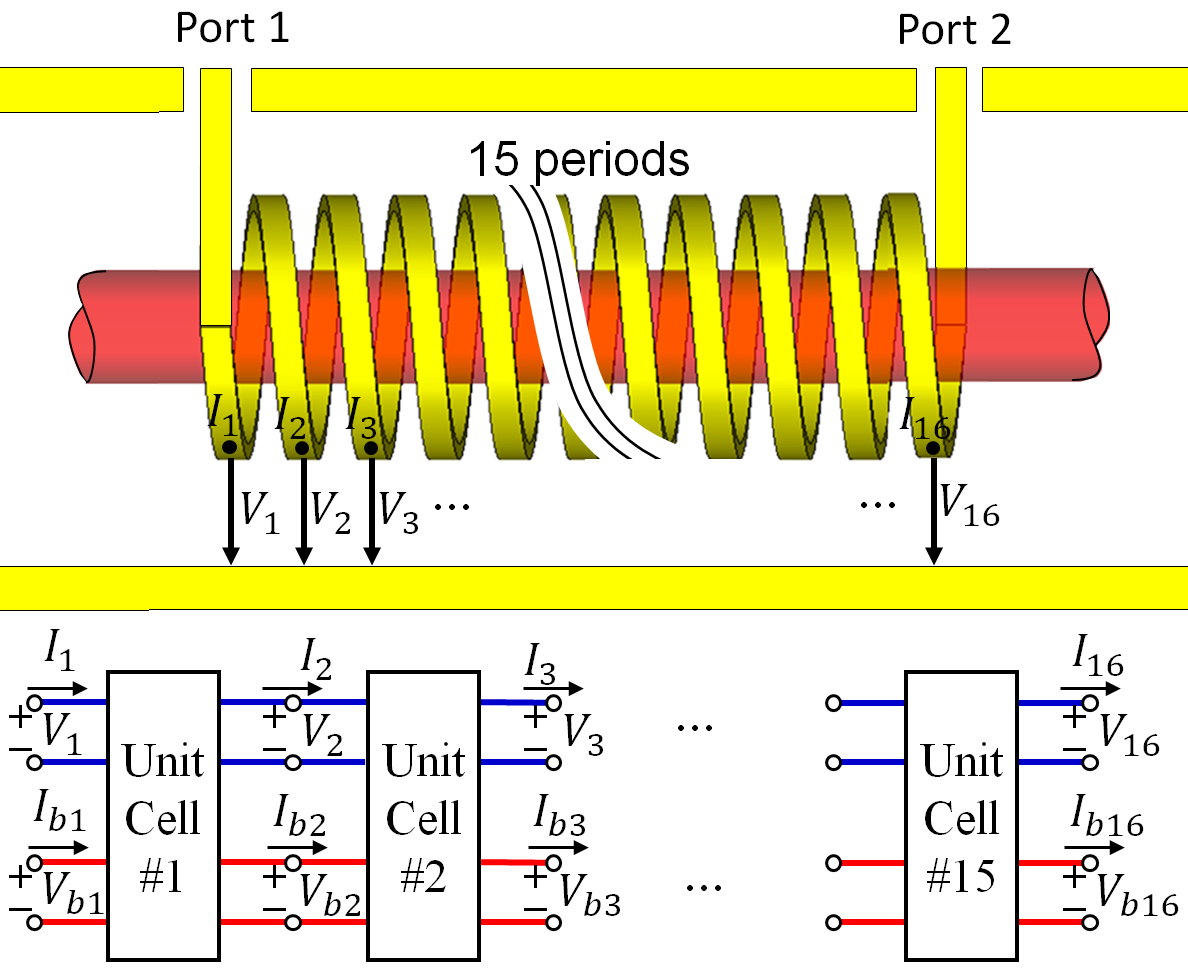}\label{Fig:TWT_Helix_Geom_b}}
\caption{(a) Geometry of the SWS of a TWT amplifier made of a circular metallic
waveguide and a copper helix with three dielectric supports. RF input
and output are through the coaxial Ports 1 and 2. (b) Circuit model
and numbering scheme used to construct the state vectors that are
used to determine the interactive system transfer matrix. Each unit
cell has four ports with equivalent voltages and currents: ($V_{n}$,
$I_{n}$) for the two EM ports and ($V_{bn}$, $I_{bn}$) for the
two electron beam ports, therefore each unit cell is described by
a $4\times4$ transfer matrix $\mathbf{T}_{u}$.\label{Fig:TWT_Helix_Geom}}
\end{figure}

We demonstrate how the technique described in the previous section
is applied considering an illustrative example made of a C-Band TWT
amplifier described in \cite{srivastava2000design} and shown in Fig.
\ref{Fig:TWT_Helix_Geom_a}. Such TWT amplifier operates at 4 GHz
and uses a solid linear electron beam with radius of 0.63 mm, with
dc kinetic voltage of 3 kV and dc current of 75 mA. The axial dc magnetic
field used to confine the electron beam is 0.6 T. The TWT uses a copper
helix tape with inner radius of 1.26 mm and period of $d=$0.76 mm,
and the metallic tape has $0.175$ mm thickness and $0.325$ mm width.
The metallic circular waveguide has radius of $3$ mm and the three
supports that physically hold the helix are made of Anisotropic Pyrolytic
Boron Nitride (APBN) with relative dielectric constant of 5.12 and
and with width of $0.7$mm. The numerical simulation consists of 15
periods of the structure. We show the numbering scheme used to define
the voltages and currents representing the EM wave and the space-charge
wave at the begin and end of each unit-cell in Fig. \ref{Fig:TWT_Helix_Geom_b}.
While the whole structure is considered in PIC simulations, when we
apply the proposed method we do not consider the first and the last
state vectors (i.e., at the beginning and end of of the first and
last unit cells) because the begin and the end of the SWS are close
to the coaxial waveguide input and output ports, and higher order
modes due to the transformation of the TEM wave in the coaxial waveguide
into the SWS modes would influence the calculation of the unit-cell
transfer matrix.

\begin{figure}
\begin{centering}
\centering \subfigure[]{\includegraphics[width=0.32\columnwidth]{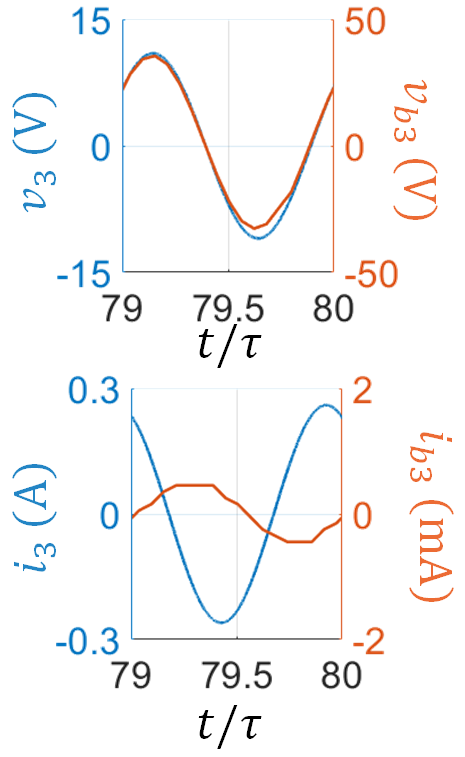}}
\subfigure[]{\includegraphics[width=0.32\columnwidth]{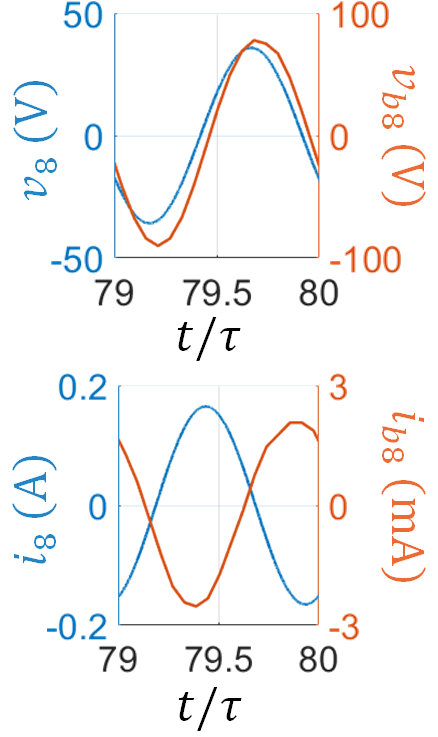}}
\subfigure[]{\includegraphics[width=0.32\columnwidth]{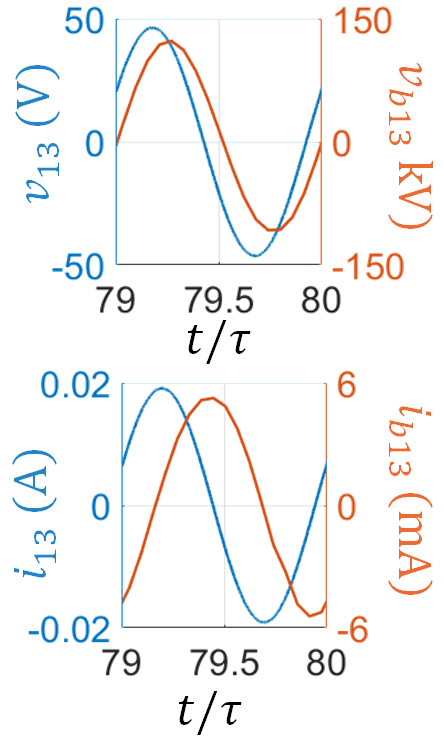}}
\par\end{centering}
\centering{}\caption{Time-domain state vector $\mathbf{\boldsymbol{\psi}}_{n}(t)$ monitored
at different $n=3,8,13$ locations of TWT at steady state: at the
entrance of the (a) $3^{rd}$, (b) $8^{th}$ and (c) $13^{th}$ unit-cell.
The four plotted quantities are provided by a full-wave simulation
based on the particle-in-cell (PIC) method. The phasor-domain representations
of the state vectors $\mathbf{\Psi}_{n}$ ($3\protect\leq n\protect\leq14$
) to be used in the proposed method to find the modal dispersion are
calculated from the time domain signals $(v_{n},i_{n})$ and $(v_{bn},i_{bn})$.
\label{Fig:Signals_Vs_Time}}
\end{figure}

First, we study the system eigenmodes at single frequency by exciting
Port 1 and Port 2 with 10 Watt and 5 Watt, receptively, at 4 GHz.
Note that indeed, even if the actual TWT operation involves feeding
from one side only, the scheme explained in the above section may
need a SWS feed from both the left and right ports to be sure we excite
all the modes. It is important to mention that changing the power
levels at Port 1 and Port 2 shall not significantly affect the resulting
eigenmode solution as long as the power levels are not too high to
involve non-linear regimes. The total number of charged particles
used to model the electron beam in the PIC simulation is about $10^{6}$
whereas the whole space in the SWS structure is modeled using $8\times10^{5}$
mesh cells. Once steady state is reached, the time-domain state vector
$\mathbf{\mathbf{\boldsymbol{\psi}}}_{n}(t)$ is monitored at each
$n^{th}$ unit-cell. Figure \ref{Fig:Signals_Vs_Time} shows the four
components (\ref{eq:TD-StateVect}) of the state vector at the $3^{rd}$,
$8^{th}$ and $13^{th}$ unit cells in the time interval between $79T$
and $80T$, where $T=0.25$ ns is the time period at 4 GHz. The phasor-domain
representation (\ref{eq:FD-StateVect}) of the state vector $\mathbf{\Psi}_{n}$
is calculated using (\ref{eq:phasor_Trans}) with $t_{ref}=79T=19.75$
ns.

\begin{figure}
\begin{centering}
\includegraphics[width=0.65\columnwidth]{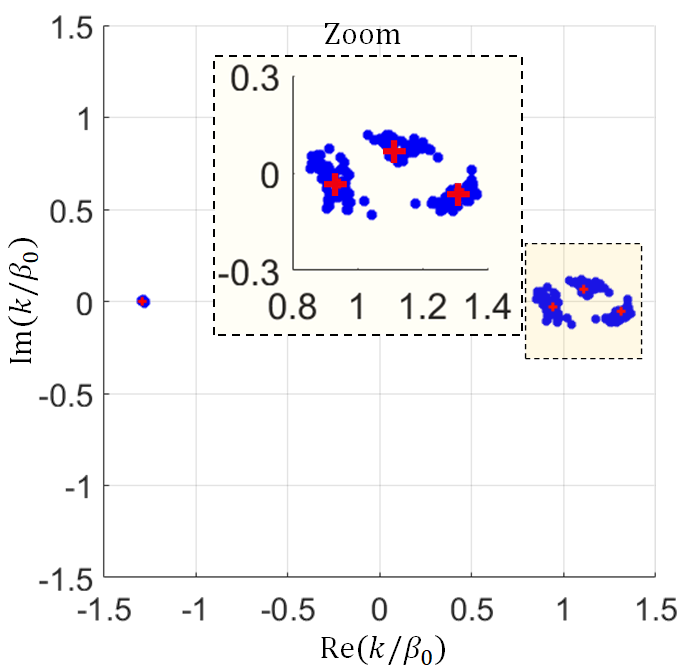}
\par\end{centering}
\centering{}\caption{Determination of the four complex wavenumbers $k_{m}$ of the four
eigenmodes ($m=1,2,3,4$) in the interactive (hot) electron beam-EM
mode in the helix-based SWS at $f=4$ GHz, when using a beam voltage
and current of $3$ kV and $75$ mA, respectively. The wavenumbers
are normalized to space-charge wave wavenumber $\beta_{0}=\omega/u_{0}$
of the beam alone, i.e., when it does not interact with the EM wave.
The red crosses represent the four wavenumbers obtained from the eigenvalues
of the transfer matrix $\mathbf{T}_{u,best}$, where $\mathbf{T}_{u,best}$
is the best-approximate solution of the overdetermined system in (\ref{eq:Tmatrix-StateVector}).
The blue dots represent different sets of four wavenumbers obtained
from the eigenmvalues of different sets of transfer matrices $\mathbf{T}_{u,qijk}$,
where $\mathbf{T}_{u,qijk}$ are the solutions obtained from (\ref{eq:small_sys})
using different combinations of indices $q,i,j$ and $k$. \label{Fig:Wavenumbe_Zplane}}
\end{figure}

The four wavenumbers of the eigenmodes of the interacting SWS system
are shown in Fig. \ref{Fig:Wavenumbe_Zplane} based on results from
Eq. (\ref{eq:Over_sys}) leading to the four red crosses, and from
Eq. (\ref{eq:small_sys}) leading to the various blue dots. The scattered
blue dots represent 75 sets of four complex wavenumbers associated
with 75 sets of transfer matrices obtained from Eq. (\ref{eq:small_sys})
using 75 combinations of $q,i,j$ and $k$ to give the highest 75
determinants of the matrix $\mathbf{w}_{1,qijk}$ out of the all 330
combinations. It is important to mention that the solutions associated
with the rest 255 combinations $q,i,j$ and $k$ were ignored because
they result in almost singular matrices $\boldsymbol{\mathrm{w}}_{1,qijk}$
and $\boldsymbol{\mathrm{w}}_{2,qijk}$.

The results in Fig. \ref{Fig:Wavenumbe_Zplane} show a good agreement
between the red crosses that represent the four wavenumbers obtained
from the eigenvalues of $\mathbf{T}_{u,best}$ where $\mathbf{T}_{u,best}$
is the best-approximate solution of the overdetermined system in (\ref{eq:Over_sys})
and the blue dots that represent the eigenmodes of various estimates
of $\mathbf{T}_{u,qijk}$ obtained from solutions of (\ref{eq:small_sys}).
It is important to point out that the small deviations between the
complex wavenumbers obtained from different solutions is due to non-idealities
resulting in having non-identical unit-cells along the structure.
This may happen due to the electron beam non-linearity and the change
of the beam kinetic voltage along the TWT, in addition to the errors
due to finite mesh and finite number of charged particles used to
model the TWT dynamics. The resulting eigenmodes are qualitatively
in good agreement with the expected description using the theoretical
transmission line-based Pierce model \cite{pierce1951waves} that
basically says that there exist three wavenumbers with positive real
part, and one of them has positive imaginary part which describes
the amplification of the EM wave along the SWS.

\begin{figure}
\begin{centering}
\includegraphics[width=1\columnwidth]{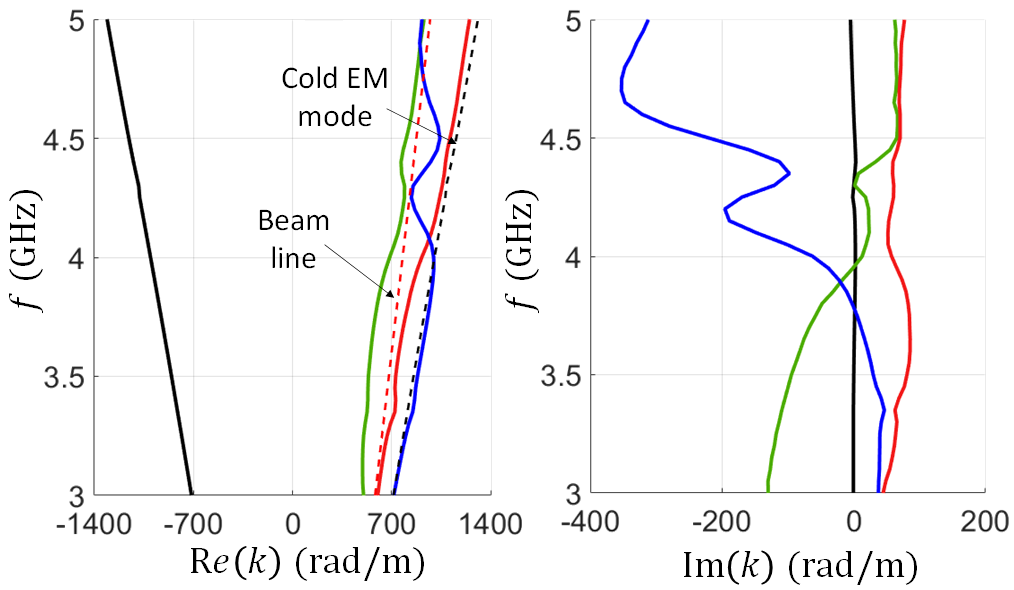}
\par\end{centering}
\centering{}\caption{Hot dispersion diagram for the four complex wavenumbers versus frequency
for the beam-EM wave interactive system made of the hot helix-based
SWS working at C-Band, using an electron beam with $3$ kV and $75$
mA. Dashed lines describe the cold case: the dashed red line represents
the wavenumber of the space-charge wave $\beta_{0}=\omega/u_{0}$
whereas the black dashed line represents the wavenumber of cold SWS;
i.e., assuming no interaction. There are three complex (interactive)
modes with positive-real part of the wavenumber, in agreement with
Pierce theory \cite{pierce1951waves}.\label{Fig:Disp_Vs_Freq_Helix}}
\end{figure}

The wavenumber-frequency dispersion describing the eigenmodes in the
hot SWS is determined by running multiple PIC simulations at different
frequencies and then determining the transfer matrix of the unit-cell
at each frequency using Eq. (\ref{eq:Over_sys}), i.e., the result
shown by the red crosses. In other words we repeat the red-cross results
shown in Fig. \ref{Fig:Wavenumbe_Zplane} at various frequencies.
The dispersion diagram of the four modes in the hot EM-electron beam
system is shown in Fig. \ref{Fig:Disp_Vs_Freq_Helix} (solid lines)
using 42 frequency points (42 PIC simulations). The dashed red and
blue lines represent the space-charge wave (i.e., the beam line) and
EM modes when they are uncoupled (i.e., in the ``cold'' case). The
dispersion of the EM mode in the cold SWS (dashed black) is found
by the finite element method-based eigenmode solver implemented in
CST Studio Suite by numerically simulating only one unit-cell of the
cold helix SWS.

\begin{figure}
\begin{centering}
\includegraphics[width=0.95\columnwidth]{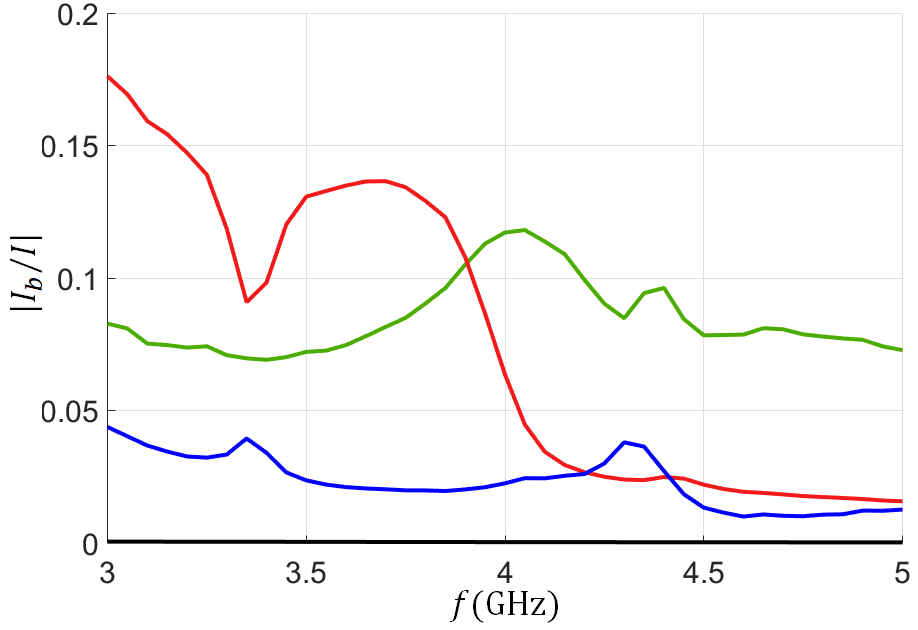}
\par\end{centering}
\centering{}\caption{Ratio of the magnitudes of the phasors representing the electron beam
ac current and EM-wave equivalent-transmission line current versus
frequency when the electron beam has adc kinetic voltage of $3$ kV
and dc current of $75$ mA. The growing mode with $\mathrm{Im}(k)$\textgreater 0
(solid red) is composed of a strong charge-wave component. The hybrid
mode with $\mathrm{Re}(k)$\textless 0 (solid-black) is mainly made
of EM field, as expected. \label{Fig:helix_I_ratio}}
\end{figure}

The black solid line in Fig. \ref{Fig:Disp_Vs_Freq_Helix} is the
hybrid mode propagating with $\mathrm{Re}(k)$\textless 0, i.e.,
in opposite direction of the beam flow and basically no beam-EM interaction
occurs, and it is made mainly of EM field as also explained later
on in Fig. \ref{Fig:helix_I_ratio}. Indeed, as compared to the cold
EM modes, the solid-black line of the hot SWS simulation is superposed
to the dispersion line of the cold EM mode propagating in the negative
$z$-direction.

The three eigenmodes with wavenumbers with positive real part (solid
red, green, blue lines) are the eigenmodes affected by the interaction
between the electron beam's space-charge wave and the EM wave propagating
in the same direction. The solid-red curve in the dispersion relation
has $\mathrm{Im}(k)$\textgreater 0 and represents the growing eigenmode
that causes the amplification of the EM wave. The imaginary part of
the growing mode starts to decreases with increasing frequency because
the difference between the speed of the space-charge wave and that
of the EM wave in the SWS (when we consider them uncoupled) increases.
It is important to mention that a single PIC simulation could also
be investigated to find the dispersion relation when using a moderately
wide-band gaussian pulse at the input ports, however it would require
excessive post processing to be able to decompose each tone behavior
and it does not fully account for steady state regime resulting from
the interaction with the electron beam that would otherwise includes
also non-linear effects.

The hybrid modes describing the EM-charge wave interaction are further
understood by looking also at their eigenvectors, besides their wavenumbers.
Eigenvectors satisfy Eq. (\ref{eq:EigenProb}) and are the same used
in the aforementioned discussed method to find the transfer matrix.
The eigenvectors components in (\ref{eq:FD-StateVect}) describe the
relative weight of the EM wave ($V$, $I$) versus the space-charge
wave $(V_{b}$, $I_{b})$, for each of the four modes $\mathbf{\Psi}^{\mathrm{mode}m}$,
with \textit{m} = 1,2,3,4. The relative strength of the EM wave and
space-charge wave in each eigenmode is determined in Fig. \ref{Fig:helix_I_ratio}
by calculating the ratio between between the ac electron beam current
and the EM mode equivalent transmission line current. The ratio $|I_{b}/I|$
for each of the four eigenmodes is shown using colors consistent with
those in Fig. \label{Fig:Disp_Vs_Freq_Helix-1}. The figure verifies
that the eigenmode with EM wave propagating with $\mathrm{Re}(k)$\textless 0
(solid-black line), in the direction opposite to the electron beam
flow does not interact with the electron beam since the beam ac current
component associated to this mode is vanishing. The growing eigenmode
($\mathrm{Im}(k)$\textgreater 0), red line, is the one responsible
of EM amplification and it shows a good EM-charge wave interaction
from 3.6 GHz to 4.2 GHz which is in agreement with the bandwidth specification
of the amplifier given in \cite{srivastava2000design}. Indeed, the
operation of TWT as an amplifier depends on having a spatially growing
mode (along the SWS) and the only mechanism to have a growing mode
is via the coupling to the electron beam that gives back power to
the EM wave after electron bunching is formed. This plot indeed shows
that the red mode is the result of synchronization and generate a
very strong electron beam. 

\begin{figure}
\begin{centering}
\subfigure[]{\includegraphics[width=1\columnwidth]{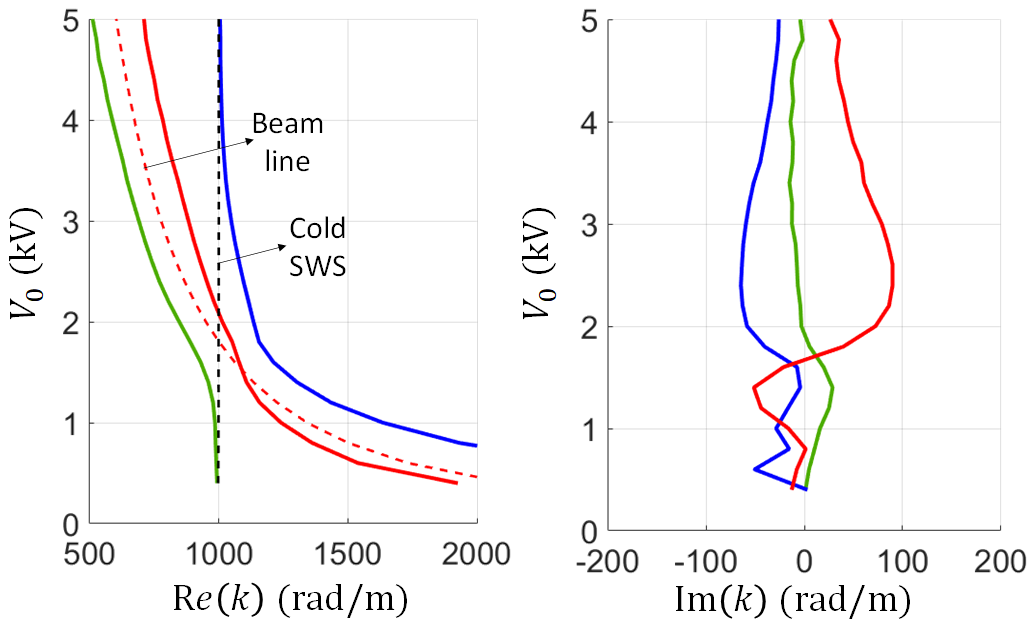}\label{Fig:Disp_vs_Volt_a}}
\par\end{centering}
\begin{centering}
\subfigure[]{\includegraphics[width=0.8\columnwidth]{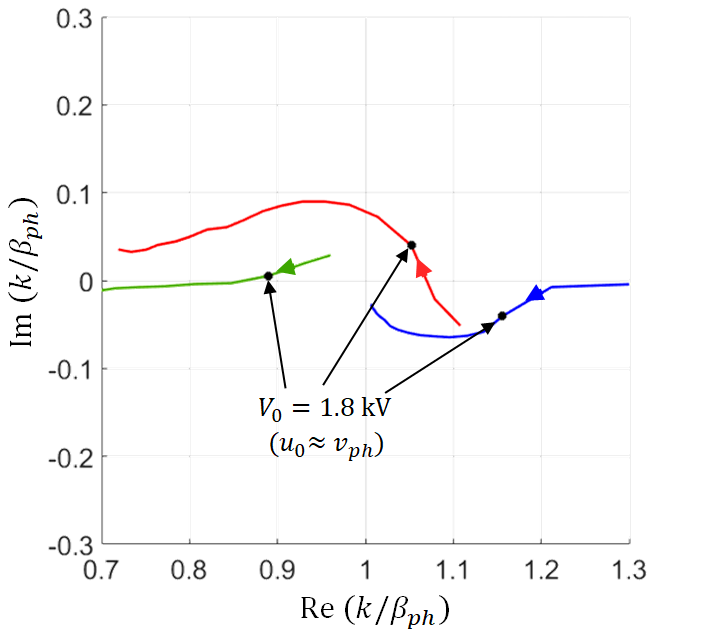}\label{Fig:Disp_vs_Volt_b}}
\par\end{centering}
\centering{}\caption{Eigenmodes complex wavenumbers for the hot helix-based SWS at frequency
$f=4$ GHz versus beam dc kinetic voltage $V_{0}$ (i.e., versus electron
velocity $u_{0}=\sqrt{2\eta V_{0}}$) at constant dc current of $75$
mA. Only the three (interactive) modes with positive-real part of
the wavenumber are shown. The amplifying mode is represented by the
red-solid curve that has $\mathrm{Im}(k)$\textgreater 0. (a) $k-V_{0}$
dispersion. The vertical dashed black line represents the wavenumber
of the EM wave in the cold SWS, whereas the dashed red curve represents
the beam space-charge wave $u_{0}=\omega/\beta_{0}$ without interaction
with the EM wave. (b) Corresponding complex plane plot for the wavenumbers
in (a), where the black dots represent the three complex modes at
the synchronization point where $u_{0}\approx v_{ph}$ at $V_{0}=1.8$
kV. The wavenumbers of modes in the hot SWS are in good agreement
with the Pierce model around the synchronization point \cite{pierce1947theoryTWT,pierce1951waves}.\label{Fig:Disp_vs_Volt}}
\end{figure}

A synchronization study is performed by observing the modes' complex
wavenumber when sweeping the beam dc voltage. In Fig. \ref{Fig:Disp_vs_Volt_a}
we show the wavenumbers for the three hot eigenmodes with positive
real part (red, green, and blue curves) when changing the beam dc
kinetic voltage $V_{0}$ while the dc beam current is maintained at
75 mA, at a constant frequency of 4 GHz. Solid lines represent the
three eigenmodes in the interactive (i.e., hot) SWS, whereas the dashed
lines represent the beam line $\beta_{0}=\omega/u_{0}$ (red dashed)
and the cold EM mode with positive wavenumber (i.e., in the cold SWS).
Since the frequency is fixed the cold EM mode has a fixed wavenumber
$\beta_{ph}=$999.8 $\mathrm{m}^{-1}$ and it is described by the
vertical black-dashed line. The phase velocity of the EM mode in the
cold SWS is $v_{ph}=\omega/\beta_{ph}=0.084c$ since $\omega=2\pi(4\times10^{9})$
rad/s, where $c=3\times10^{8}$m/s is the speed of waves in free space.
Synchronization occurs approximately in the region where the two dashed
lines, beam line and EM-wave line, have the same wavenumber, i.e.,
the same phase velocity, which happens just above 1.8 kV. The TWT
amplification factor, mainly represented by the positive imaginary
part of the red eigenmode, shows a peaking around synchronization
point of $V_{0}=$1.8 kV which is corresponding to a beam speed $u_{0}=\sqrt{2\eta V_{0}}=0.084c$,
where $\eta=e/m$ is the charge to mass ratio of the electron.

At lower or much higher dc voltage with respect to the synchronization
point would make the beam-EM interaction is weak. This is also understood
by noting that the gain starts to decay away from the interaction
voltage region and the hot modes (resulting from the interaction)
tend to overlap with the cold modes (without interaction) away from
such region. In particular at low frequency the solid green curve
tends to overlap with the cold EM mode, whereas at high frequency
the solid blue line tends to overlap with the cold EM mode. The two
solid red and blue curves at low dc voltage tend to overlap with the
beam curve (red dashed), whereas the two solid red and green curves
at high voltage tend to overlap with the beam line (red dashed): therefore
these hot modes tend to overlap with the beam line (non-interactive
beam) at high and low voltage despite some shifts which may be due
to space charge effects. Near the synchronization point, the splitting
of the solid green and blue curve (versus the red curve that does
not split) represent the strong interaction between the electron beam
and the EM field.

In Fig. \ref{Fig:Disp_vs_Volt_b} we show the complex plane mapping
of the wavenumbers for the three hot eigenmodes with $\mathrm{Re}(k)>0$,
shown in Fig. \ref{Fig:Disp_vs_Volt_a}, varying the electron beam
dc voltage $V_{0}$ (i.e., the speed of the electrons $u_{0}$). The
black dots represent the interactive modes wavenumbers when the beam
dc voltage is $V_{0}=1.8$ kV which results in an electron beam with
average speed very close to the EM mode in the cold SWS, i.e., close
to the synchronization point $u_{0}\approx v_{ph}$. The complex wavenumber
location of the three interactive modes shown in Fig. \ref{Fig:Disp_vs_Volt_b}
is in agreement with the three-wave theory of the Pierce model \cite{pierce1947theoryTWT,pierce1951waves}
around the synchronization point. When $u_{0}\approx v_{ph}$, there
are three modes with $\mathrm{Re}(k)$\textgreater 0: two of them
are waves that are slower than the electron beam average speed $u_{0}\approx v_{ph}$,
and among these two, one wave is growing in the beam direction while
the other one is decaying. The third mode is basically an unattenuated
wave that travels faster than the beam-average speed $u_{0}\approx v_{ph}$.

\begin{figure}
\begin{centering}
\includegraphics[width=0.9\columnwidth]{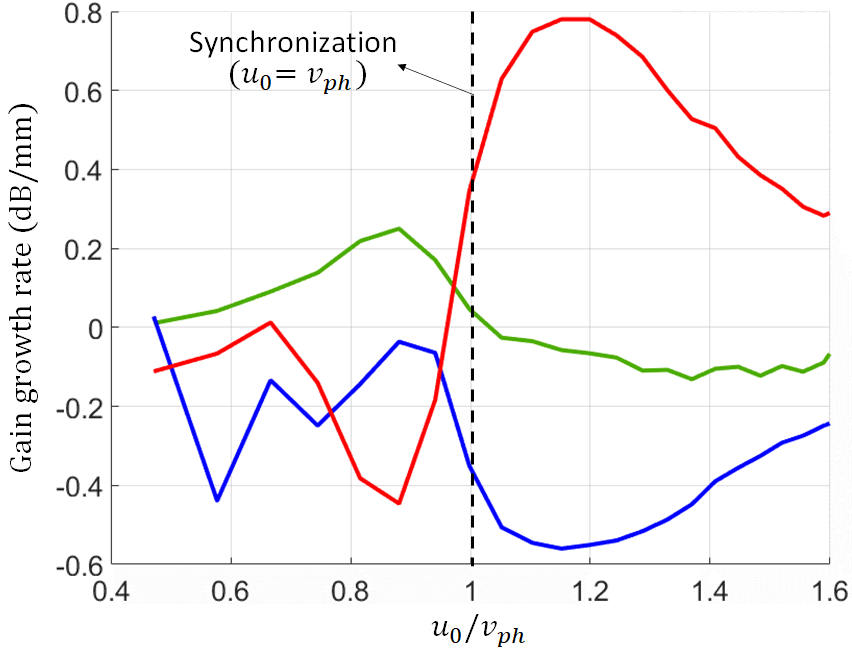}
\par\end{centering}
\centering{}\caption{Gain growth rate defined for the three interactive modes in the hot
helix-based SWS versus the electron beam average speed $u_{0}$ normalized
to the cold SWS phase velocity. We change the electron beam speed
$u_{0}$ through varying the beam dc kinetic voltage $V_{0}$ ($u_{0}=\sqrt{2\eta V_{0}}$).
As in Fig. \ref{Fig:Disp_vs_Volt}, the beam dc current is $75$ mA
and $f=4$ GHz, hence the cold EM mode phase velocity is constant
$v_{ph}=0.084c$ . The amplifying mode is represented by the red-solid
curve. The point where the gain growth rate is maximum is close to
the synchronization point when $u_{0}\approx1.2v_{ph}.$\label{Fig:Gain_growth}}
\end{figure}

The power flow for each mode of the hot (i.e. interactive) SWS is
written in the form of $P_{m}(z)=P_{0,m}e^{2\mathrm{Im}(k_{m})z}$
\cite{pozar}, where $P_{0,m}$ is the initial amount of power carried
by the same mode at $z=0$. We define the power gain resulting from
the interaction between the EM and space-charge wave for as $G_{p,m}(z)=P_{m}(z)/P_{0,m}=e^{2\mathrm{Im}(k_{m})z}.$
Thus gain growth rate for each mode is $20\log(e)\mathrm{Im}(k_{m})$
dB/m. In Figure \ref{Fig:Gain_growth} we show the gain growth rate
for the three interactive modes (those with $\mathrm{Re}(k_{m})>0$)
versus electron-beam charges average speed $u_{0}=\sqrt{2\eta V_{0}}$
(normalized by the cold SWS phase velocity $v_{ph}=\omega/\beta_{ph}$
, where $\beta_{ph}$ is the wavenumber of the EM mode in the cold
SWS). The figure shows that there is an optimal point for the interaction
(where $u_{0}/v_{ph}\approx1.18$) which is very close to synchronization
point where the power transfer from the kinetic energy of the electron
beam into RF power in the SWS is maximum, as discussed in \cite{kosmahl1984twt,pierce1947theory}.
The figure also show that the largest power transfer, from the kinetic
energy of the electron beam into RF power, occurs when the average
speed of electron beam charges is roughly above the speed of the cold
EM wave in the SWS.

In Appendix A we show the effect of changing the voltage definition
and the SWS length on the eigenmode calculations for the same helix
SWS considered in this section. There, we define the voltages representing
the EM waves as the potential differences between each two successive
helix loops. Results are qualitatively in agreement with what described
here, and only small quantitative differences are observed.

\section{Application to serpentine-based TWT amplifier}

\begin{figure}
\begin{centering}
\subfigure[]{\includegraphics[width=0.9\columnwidth]{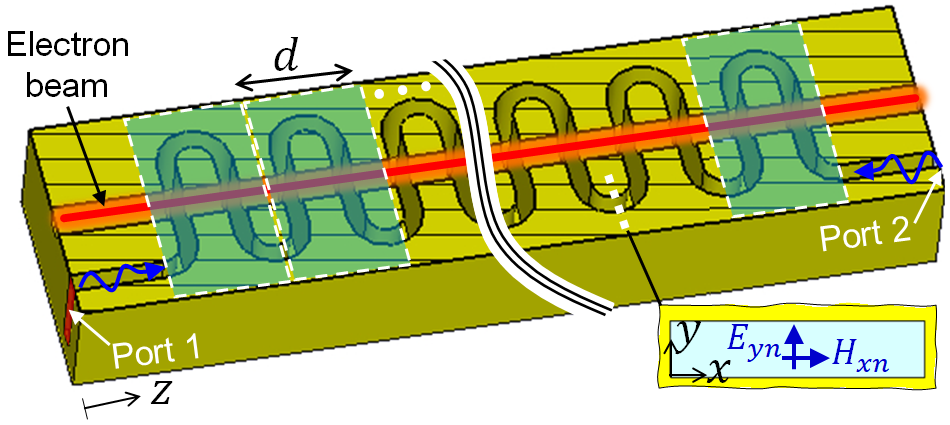}\label{Fig:Serpentine_a}}
\par\end{centering}
\begin{centering}
\subfigure[]{\includegraphics[width=0.75\columnwidth]{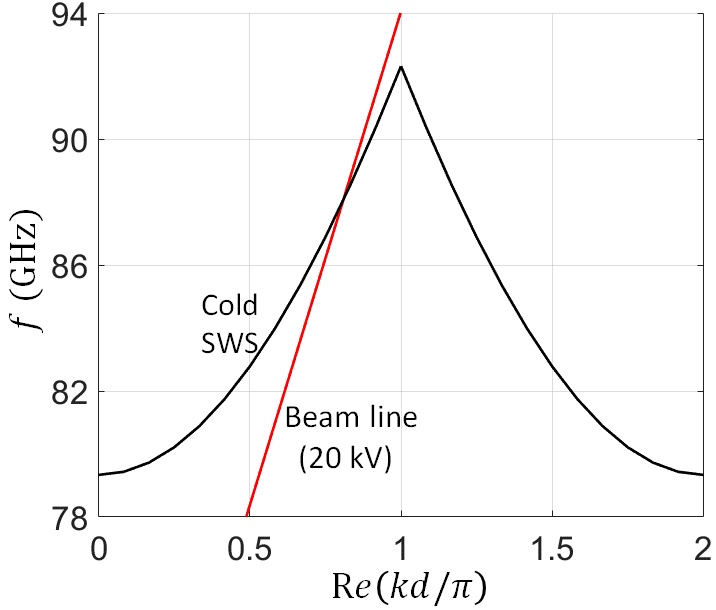}\label{Fig:Serpentine_b}}
\par\end{centering}
\centering{}\caption{TWT made of a serpentine waveguide operating at millimeter waves.
(a) Setup used to determine the dispersion relation of the eigenmodes
in the hot serpentine SWS based on PIC simulations. The unit cell
has period \textit{d} and two wave ports are located at the two terminations
of the serpentine waveguide. The inset shows the electric and magnetic
fields in the metallic rectangular waveguide. (b) Dispersion relation
showing the \textit{cold} eigenmodes of the EM wave (black) and the
beam line (red). The electron beam has adc voltage of 20 kV and it
interacts with a forward EM mode leading to TWT amplification.\label{Fig:Serpentine}}
\end{figure}

We demonstrate the utility of the proposed eigenmode solver method
to find the eigenmodes in the hot serpentine SWS operating at a millimeter
wave band. The interaction with the electron beam is periodic and
not uniform as for the case of the helix SWS. Serpentine SWSs have
recently gained a lot of interest due to the growing importance of
millimeter wave and terahertz frequencies in modern applications and
also due to the advancement of fabrication technologies such as LIGA
(LIthographie, Galvanoformung, Abformung). As an illustrative example,
we use the same geometry of serpentine SWS discussed in \cite{feng2011study,srivastava2018design},
as shown in \ref{Fig:Serpentine_a}. The serpentine waveguide is made
of copper and has rectangular cross-section of dimensions $a=1.9$
mm and $b=0.325$ mm, bending radius of $0.325$ mm (radius at half
way between inner and outer radii), straight section length of 0.6
mm, beam tunneling radius of $0.175$ mm. The TWT comprises 13 unit-cells
. An electron beam with dc voltage $V_{0}=20$ kV is used such that
the synchronization occurs with a forward EM wave leading to amplification.
In Fig \ref{Fig:Serpentine_b} we show the beam line $\beta_{0}=\omega/u_{0}$
(red line) where $u_{0}=\sqrt{2\eta V_{0}}=0.28c$, and the dispersion
diagram of EM modes in the \textit{cold} SWS (black line). Synchronization,
$v_{ph}\approx u_{0}$, between the electron beam and the forward
EM wave occurs at frequencies centered at $f=88$ GHz.

Figure \ref{Fig:Serpentine_a} shows the setup used for PIC simulations.
We consider the electron beam to have a radius of 0.13 mm, a dc current
of $0.1$ A and an axial confinement dc magnetic field of 0.6 T. Our
goals is to obtaining the dispersion of the hot serpentine SWS, i.e.,
the complex wavenumber of the hybrid modes that account for the interaction
between the electron beam and the EM wave. In our method we excite
the SWS from Port 1 with 10 Watts and from Port 2 with 5 Watts. An
amplifier has the input at one port and the output at the other one,
but here we want to excite the supported eigenmodes sufficiently to
be observed in the calculations. The serpentine supports the $\mathrm{TE_{10}}$
mode which is the only one propagating in the rectangular waveguide.
Since the serpentine waveguide does not support a TEM mode, voltage
and current cannot be uniquely defined. We use the equivalent representation
in \cite{marcuvitz1951waveguide,collin1990field,felsen1994radiation}
that models the waveguide as a transmission line with equivalent voltage
and current. Following the derivations in \cite{marcuvitz1951waveguide,collin1990field,felsen1994radiation}
for the $\mathrm{TE_{10}}$ mode in a rectangular waveguide, the transverse
fields are written as

\begin{equation}
\begin{array}{c}
E_{y}(x,y,z)=V(z)\sqrt{\dfrac{2}{ab}}\sin\left(\dfrac{\pi x}{a}\right),\\
H_{x}(x,y,z)=I(z)\sqrt{\dfrac{2}{ab}}\sin\left(\dfrac{\pi x}{a}\right).
\end{array}\label{eq:V_I_Def}
\end{equation}
Using (\ref{eq:V_I_Def}), the discrete voltages and the currents
that represent the EM state at different rectangular cross-sections
of the serpentine waveguide are found as

\begin{equation}
\begin{array}{c}
V_{n}=\sqrt{\dfrac{ab}{2}}E_{yn},\\
I_{n}=\sqrt{\dfrac{ab}{2}}H_{xn},
\end{array}
\end{equation}
where $E_{yn}$ and $H_{xn}$ are the traverse electric and magnetic
fields calculated at the center of the rectangular ($x=a/2$ and $y=b/2$)
waveguide cross section as shown in the inset in Fig. \ref{Fig:Serpentine_a}
and they are calculated at the unit cells boundaries shown in Fig.
\ref{Fig:Serpentine_a}.

\begin{figure}
\begin{centering}
\subfigure[]{\includegraphics[width=0.65\columnwidth]{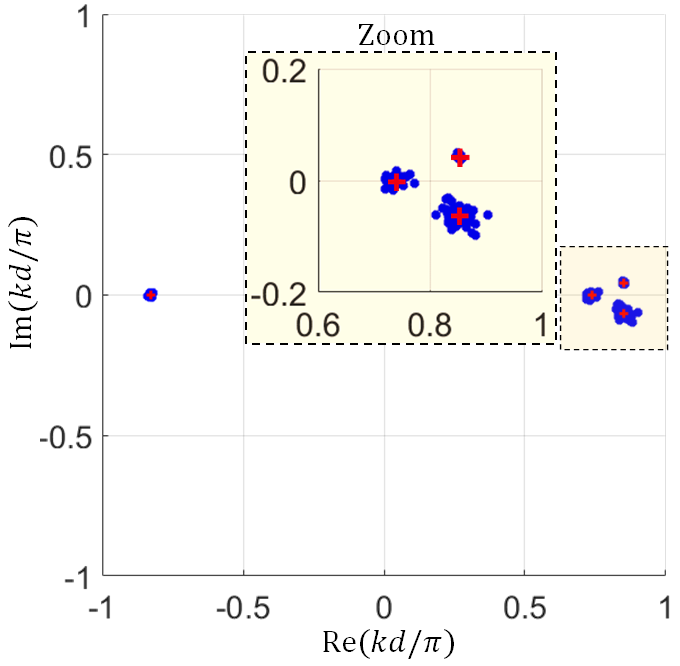}\label{Fig:Serpentine_2_a}}
\par\end{centering}
\begin{centering}
\subfigure[]{\includegraphics[width=1\columnwidth]{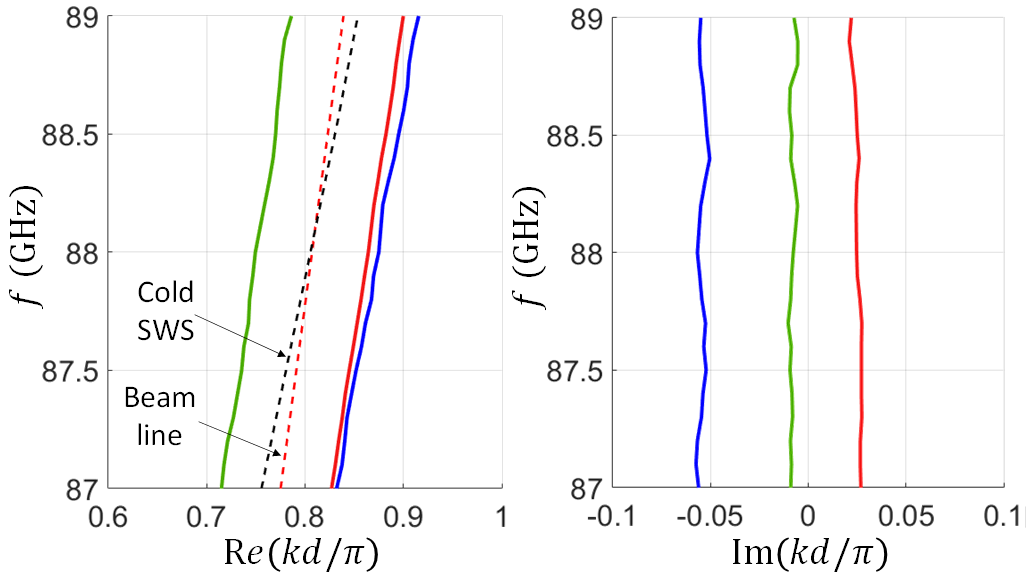}\label{Fig:Serpentine_2_b}}
\par\end{centering}
\centering{}\caption{Complex wavenumbers of the eigenmodes in the interactive (hot) electron
beam-EM mode in the serpentine SWS in Fig. \ref{Fig:Serpentine}(a),
evaluated using PIC simulations, assuming a beam voltage of $20$
kV and current of $0.1$ A. (a) Complex plane plot of the four complex
wavenumbers in the first Brillouin zone, at $f=88$ GHz. The red crosses
represent the four wavenumbers obtained from the transfer matrix $\mathbf{T}_{u,best}$
obtained from (\ref{eq:Over_sys}). Blue dots represent different
sets of four wavenumbers obtained from different sets of transfer
matrices $\mathbf{T}_{u,qijk}$ obtained from (\ref{eq:small_sys})
using different combinations of indices $q,i,j$ and $k$. Both results
are in agreement with the Pierce model \cite{pierce1947theoryTWT,pierce1951waves}.
(b) Complex wavenumber-frequency dispersion relations (solid lines).
The dashed-red line is the wavenumber dispersion of the electron beam's
space-charge wave described by $\beta_{0}=\omega/u_{0}$, whereas
the dashed-black line represents the wavenumber of the EM mode in
the cold SWS, i.e., assuming no beam-EM interaction. The figures show
the three hybrid modes around the synchronization point, i.e., the
three modes with wavenumber with positive real part. The mode associated
to the solid-red curve is responsible for amplification since $\mathrm{Im}(k)$\textgreater 0.\label{Fig:Serpentine_2}}
\end{figure}

We start by studying the eigenmode wavenumbers in the interacting
SWS system at constant frequency $f=88$ GHz, which is very close
to the synchronization point where $u_{0}=v_{ph}$. The four complex
wavenumbers of the hybrid modes are shown in Fig. \ref{Fig:Serpentine_2_a}
based on results from Eq. (\ref{eq:Over_sys}) leading to the four
red crosses, and from Eq. (\ref{eq:small_sys}) leading to various
blue dots. The scattered blue dots represent 37 sets of four complex
wavenumbers associated the largest 37 determinants of the matrix $\mathbf{w}_{1,qijk}$
out of the all 126 combinations. The blue dots cluster around the
four red crosses, as expected. It is important to mention that we
ignored the state vectors of the first and the last two unit cells
to generate our results because they may involve high order modes
that affect the results. The complex wavenumbers locations of the
three interactive modes (those with $\mathrm{Re}(k)$\textgreater 0)
shown in Fig. \ref{Fig:Serpentine_2_a} are in good agreement with
predictions of the three-wave theory of the Pierce model \cite{pierce1947theoryTWT,pierce1951waves}.

In Fig. \ref{Fig:Serpentine_2_b} we show the modal dispersion relation
accounting for the EM-beam interaction in the serpentine SWS using
21 frequency points. The hot dispersion diagrams are obtained using
the best-approximate solution $\mathbf{T}_{\mathit{u,best}}$ of the
overdetermined system in (\ref{eq:Over_sys}). As in the previous
section, dashed lines represent the two uncoupled systems: the beam
line (dashed red) and the EM mode in the cold SWS (dashed black).
The electron beam has a dc voltage of $V_{0}=20$ kV so the electron
beam interacts with the forward EM wave resulting in an eigenmode
with positive imaginary part leading to TWT amplification (red curve).
We show only the three modes with positive real wavenumber. One wavenumber
has positive imaginary part (solid red curve), which is responsible
for amplification, whereas the other two modes resulting from the
interaction are decaying and unattenuated modes, in agreement with
the Pierce model \cite{pierce1951waves}. The gain per unit-cell associated
to the $m^{th}$ mode is defined as

\begin{equation}
G_{p,m}=\frac{P_{m}[(n+1)d]}{P_{m}(nd)}=e^{2\mathrm{Im}(k_{m})d},
\end{equation}
which is equivalent to $20\log(e)\mathrm{Im}(k_{m}d)$ dB. The imaginary
part of the wavenumber of the amplification mode (solid red) is almost
constant and equal to$\mathrm{Im}(kd)\approx0.027\pi$ in the frequency
range from 87 GHz to 89 GHz shown in Fig. \ref{Fig:Serpentine_2_b},
because the phase synchronization condition is almost satisfied for
the considered frequency band, i.e., $\beta_{ph}\approx\beta_{0}$
over the shown band (relative band of $2.2\%$ around $88$GHz) in
this example. Thus, the gain per period resulting from the amplification
mode (solid red) is $20\log(e)\times0.027\pi\approx$ 0.74 dB in this
frequency range which is close to the small-signal gain 1 dB reported
in \cite{srivastava2018design} that was obtained by simulating the
serpentine TWT amplifier at 90 GHz.

\begin{figure}
\begin{centering}
f\subfigure[]{\includegraphics[width=0.65\columnwidth]{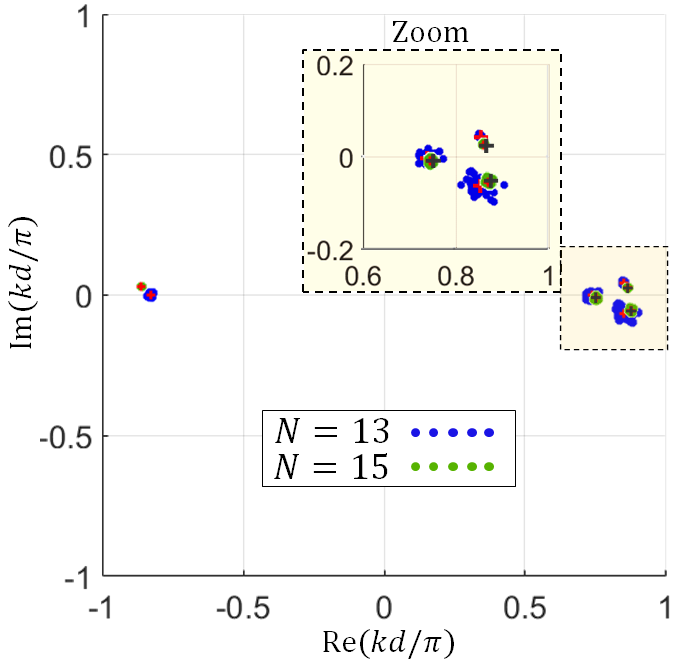}\label{Fig:Repeat_a}}
\par\end{centering}
\begin{centering}
\subfigure[]{\includegraphics[width=1\columnwidth]{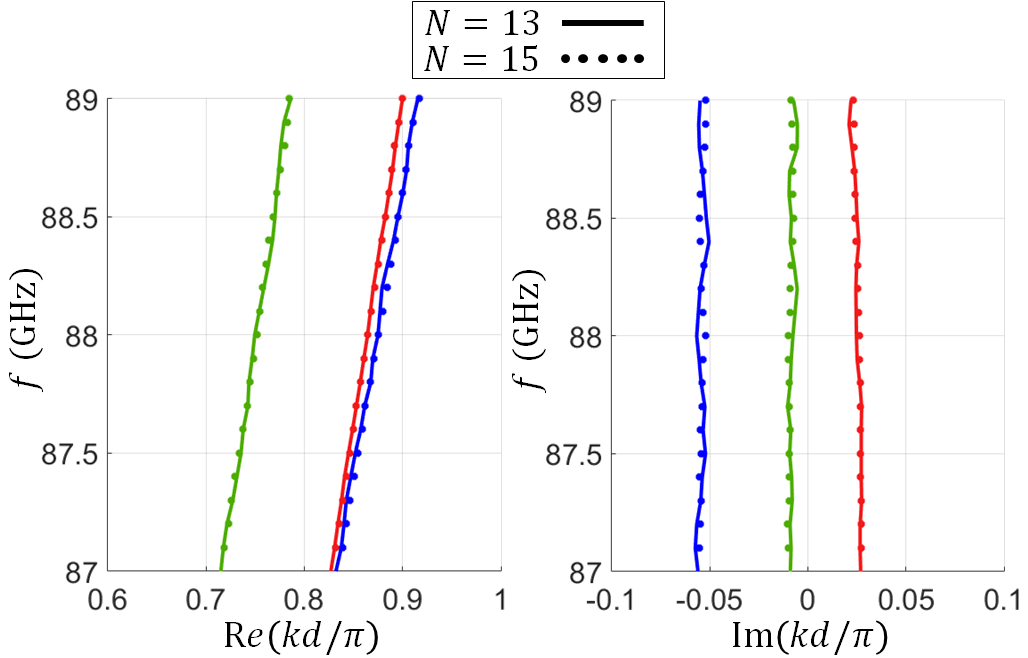}\label{Fig:Repeat_b}}
\par\end{centering}
\centering{}\caption{Repeatability test for obtaining the complex wavenumbers of the eigenmodes
in the interactive (hot) electron beam-EM wave system in Fig. \ref{Fig:Serpentine}(a),
evaluated using data from PIC simulations of finite-length serpentine
SWS structures with $N=13$ and $N=15$ unit cells. (a) Complex plane
plot of the four complex wavenumbers in the first Brillouin zone,
at $f=88$ GHz. Blue and green dots represent the wavenumbers obtained
from the distinct determined solutions using $N=13$ and $N=15$ unit
cells, respectively. Whereas the red and black crosses represent the
wavenumbers obtained from the best-approximate solution of the overdetermined
system using $N=13$ and $N=15$ unit cells, respectively. (b) Complex
wavenumber-frequency dispersion relations obtained from the best-approximate
solution $\mathbf{T}_{u,best}$of the overdetermined system (\ref{eq:Over_sys})
(red and black crosses in (a)). Solid and dotted curves represent
the dispersion obtained from the simulation of the finite-length SWS
with $N=13$ and $N=15$ unit cells, respectively.}
\end{figure}

A repeatability study is performed by obtaining the dispersion relation
for the serpentine hot SWS using two different numbers of unit cells
(all other parameters are kept the same for the two TWT configurations
as previously described). Indeed, the transfer matrix of the hot SWS
unit-cell $\mathbf{T}_{u}$ should not be a function of the used number
of unit-cells, therefore, the dispersion relation obtained using different
numbers of unit cells should not be affected. In Fig. \ref{Fig:Repeat_a}
we show the complex wavenumberscalculated from the unit-cell transfer
matrix $\mathbf{T}_{u}$ estimated using data from PIC simulations
of two SWSs with $N=13$ (as in the previous example) and $N=15$
unit cells. In each case, we plot 37 sets of four wavenumbers obtained
from 37 estimated transfer matrices (distinct determined solutions):
blue and green dots represent the cases with $N=13$ and $N=15$ unit
cells, respectively. The plotted sets are the ones associated with
the highest 37 determinants of the matrices $\mathbf{w}_{1,qijk}$
used to determine the unit-cell transfer matrices $\mathbf{T}_{u,qijk}$
out of the all 126 combinations for the case with $N=13$ unit cells
and 330 combinations for the case with $N=15$ unit cells . The red
and black crosses represent the wavenumbers obtained from the best-approximate
solution of the overdetermined system for the cases with $N=13$ and
$N=15$ unit cells, respectively. The clustering of wavenumbers at
88 GHz in Fig. \ref{Fig:Repeat_a} shows a good agreement between
the two cases based on $N=13$ and $N=15$ unit cells. In Fig. \ref{Fig:Repeat_b}
we compare the dispersion relation obtained based on the best-approximate
solution of the overdetermined system (red and black crosses in Fig.
\ref{Fig:Repeat_a}): solid and dotted curves represent the dispersion
obtained from simulating the hot SWS with $N=13$ and $N=15$ unit
cells, respectively. The figure shows an almost negligible discrepancy
between the dispersion diagrams obtained using two SWS lengths.

\section{Conclusion}

A TWT eigenmode solver to determine the complex wavenumbers of the
hybrid eigenmodes in hot SWSs (i.e., accounting for the interaction
of the electron beam and the EM wave) has been demonstrated using
a novel approximate technique based on data obtained from PIC simulations.
The technique has been able to predict the growing mode's complex
wavenumber in both TWTs studied here based on: (i) a SWS made of a
circular waveguide with a helix working in the GHz range; (ii) a SWS
made of a serpentine waveguide for millimeter wave amplification.
The method is based on elaborating the data obtained from time domain
PIC simulations, hence accounting for the precise waveguide geometry,
materials, losses, and beam space charge effects. We believe that
the proposed technique is a powerful tool for the understanding and
the design of TWTs amplifiers.

In the present investigation we have focused on determining the complex
wavenumbers of the hybrid EM-beam modes, but the same technique can
also be used to estimate the performance of longer TWTs based on simulations
of shorter TWTs, i.e., it can be used in the design and initial optimization
of TWTs. The determination of wavenumbers and eigenvectors of all
the hybrid modes supported in a TWT amplifiers discussed here can
also be useful to study mode degeneracy conditions in hot SWSs as
those investigated in \cite{mealy2019backward,mealy2019exceptional,mealy2020exceptional}.

\section{Acknowledgment}

This material is based upon work supported by the Air Force Office
of Scientific Research award number FA9550-18-1-0355 and award number
FA9550-20-1-0409. The authors are thankful to DS SIMULIA for providing
CST Studio Suite that was instrumental in this study.

\appendices{}

\section{Helix dispersion relation using an alternative voltage definition
and different structure lengths}

In this appendix we test how changing the voltage definition impacts
the eigenmodes calculations for the TWT made of a helix SWS considered
in Sec. \ref{sec:Helix_TWT}. We also test how changing the length
of the simulated TWT impacts the eigenmodes calculations.

\subsection{Changing voltage definition}

\begin{figure}
\begin{centering}
\subfigure[]{\includegraphics[width=0.8\columnwidth]{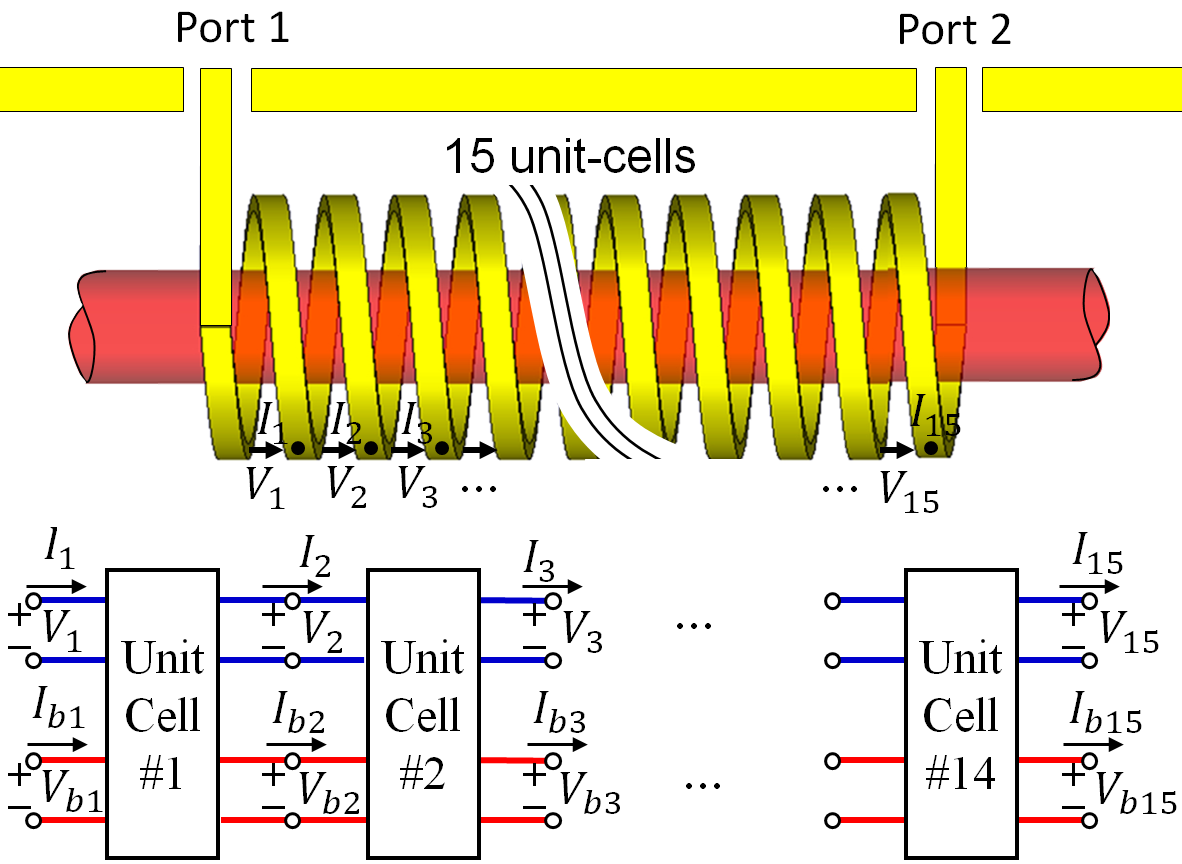}\label{Fig:volt_def_a}}
\par\end{centering}
\begin{centering}
\subfigure[]{\includegraphics[width=1\columnwidth]{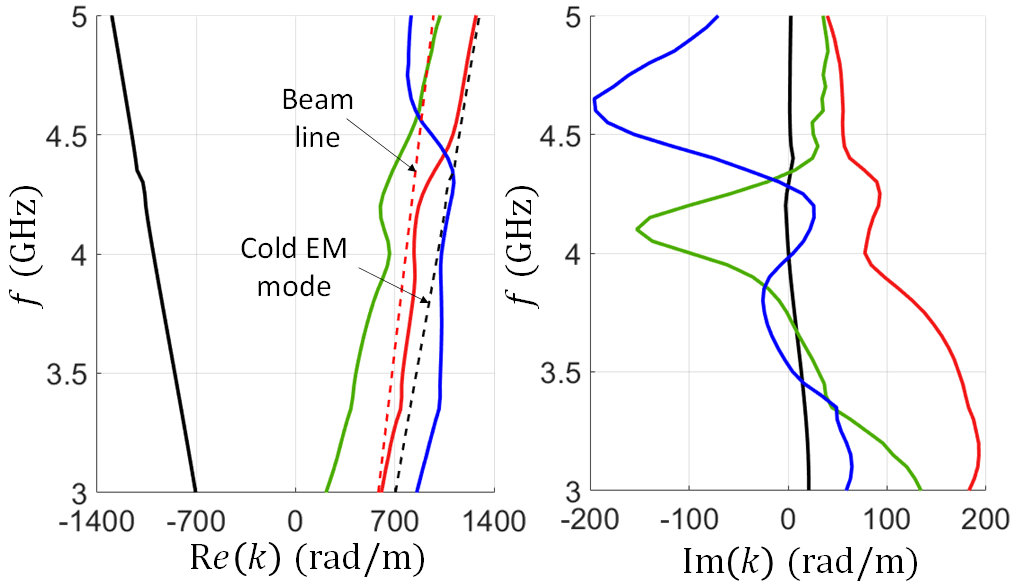}\label{Fig:volt_def_b}}
\par\end{centering}
\begin{centering}
\subfigure[]{\includegraphics[width=1\columnwidth]{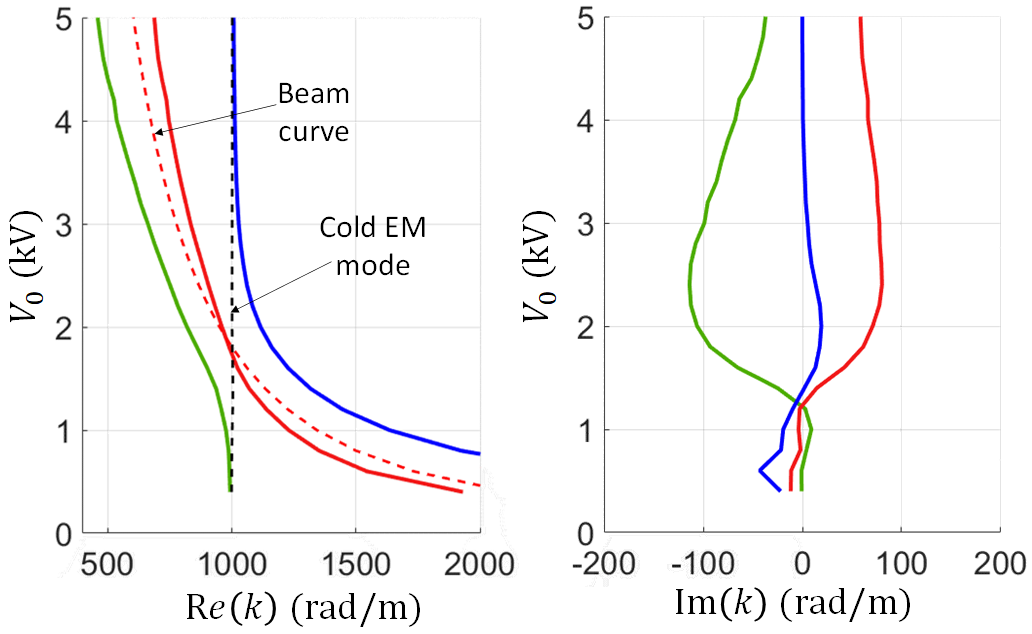}\label{Fig:volt_def_c}}
\par\end{centering}
\centering{}\caption{(a) Setup used to determine the frequency-wavenumber dispersion relation
assuming the EM waves to be represented using voltages between each
two successive loops. Hot dispersion diagram for the four complex
wavenumbers versus: (b) frequency using an electron beam with $3$
kV and $75$ mA, and (c) electron beam dc kinetic voltage at constant
frequency $f=4$ GHz and dc current of $75$ mA.\label{Fig:volt_def}}
\end{figure}

The structure we study here has the same helix parameters and length
as the one presented in Sec. \ref{sec:Helix_TWT}. The only difference
is that here we define the voltages representing the EM field as the
potential differences between each two consecutive helix loops. We
show in Fig. \ref{Fig:volt_def_a} the definition and the numbering
scheme for voltages and currents representing the EM and space-charge
waves,that are used to construct the circuit network model of each
unit-cell. Note that although the structure has 15 periods but it
is modeled using 14 network unit-cells because the new voltage definition
involve two loops of the helix as shown in Fig. \ref{Fig:volt_def_a}.

Theoretically, changing the voltage definition should not impact the
eigenvalues, i.e., the determination of the hot SWS modal wavenumbers,
however, it affects the calculation of the eigenvectors of the system.
We show in Fig. \ref{Fig:volt_def_b} and Fig. \ref{Fig:volt_def_c}
the the four complex modal wavenumbers of the hot EM-electron beam
system, versus frequency and beam voltage, respectively. The small
discrepancy between the wavenumber obtained in Sec. \ref{sec:Helix_TWT}
and the one obtained here based on a different voltage definition
may be explained considering the electron beam non-linearity and the
change of the beam kinetic voltage along the TWT, in addition to the
errors due to finite mesh and finite number of charged particles used
to model the TWT dynamics.

\subsection{Changing the SWS length}

\begin{figure}
\begin{centering}
\includegraphics[width=0.65\columnwidth]{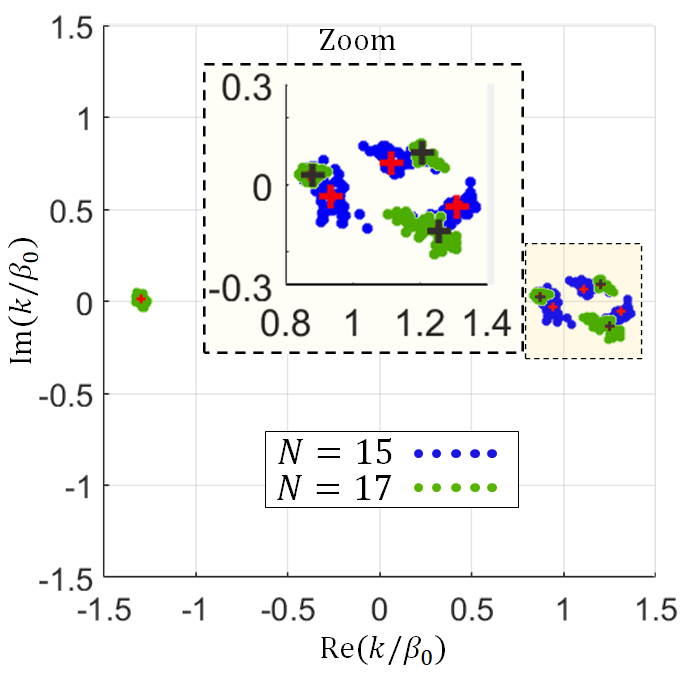}
\par\end{centering}
\centering{}\caption{Repeatability test for obtaining the complex wavenumbers of the eigenmodes
in the interactive (hot) helix SWS evaluated using data from PIC simulations
for finite structures with number of unit-cells $N=15$ (blue dots)
and $N=17$ (green dots) at $f=4$ GHz. \label{Fig:Wavenumbe_Zplane_Repeat}.
The blue and green dots represent the wavenumbers obtained from the
distinct determined solutions using $N=15$ and $N=17$ unit cells,
respectively. Whereas, the red and black crosses represent the wavenumbers
obtained from the best-approximate solution of the overdetermined
system using $N=15$ and $N=17$ unit cells, respectively.}
\end{figure}

We study the effect of changing the simulated TWT length on the calculation
of the four complex wavenumbers calculated from the transfer matrix
$\mathbf{T}_{u}$ estimated using data from PIC simulations of two
SWSs with $N=15$ and $N=17$ unit cells. All the used parameters
here are kept the same as the ones in Sec. \ref{sec:Helix_TWT} except
that we change the number of the helix SWS unit cells to be $N=17$,
instead of $N=15$. In Fig. \ref{Fig:Wavenumbe_Zplane_Repeat} we
show the complex wavenumbers. In each case, we plot 75 sets of four
wavenumbers obtained from 75 estimated transfer matrices (distinct
determined solutions):blue and green dots represent the cases with
$N=15$ and $N=17$ unit cells, respectively. The plotted sets are
the ones associated with the highest 75 determinants of the matrices
$\mathbf{w}_{1,qijk}$ used to determine the unit-cell transfer matrices
$\mathbf{T}_{u,qijk}$ out of the all 330 and 715 combinations for
the case with $N=15$ and $N=17$ unit cells, respectively. Red and
black crosses represent the wavenumbers obtained from the best-approximate
solution of the overdetermined system for the cases with $N=13$ and
$N=15$ unit cells, respectively. The clustering around distinct wavenumbers
is consistent when considering the two SWS lengths. The small but
noticeable discrepancy in values between the wavenumbers clustering
for both cases may be explained as the result of the electron beam
non-linearity and other factors like the electron beam loss of energy
along the SWS.

\bibliographystyle{ieeetr}
\bibliography{myref}

\begin{thebibliography}{10}

\bibitem{minenna2019traveling}
D.~F. Minenna, F.~Andr{\'e}, Y.~Elskens, J.-F. Auboin, F.~Doveil, J.~Puech, and
  {\'E}.~Duverdier, ``The traveling-wave tube in the history of
  telecommunication,'' {\em The European Physical Journal H}, vol.~44, no.~1,
  pp.~1--36, 2019.

\bibitem{li2018design}
X.~Li, X.~Huang, S.~Mathisen, R.~Letizia, and C.~Paoloni, ``Design of 71--76
  $\text{GHz}$ double-corrugated waveguide traveling-wave tube for satellite
  downlink,'' {\em IEEE Transactions on Electron Devices}, vol.~65, no.~6,
  pp.~2195--2200, 2018.

\bibitem{booske2008plasma}
J.~H. Booske, ``Plasma physics and related challenges of
  millimeter-wave-to-terahertz and high power microwave generation,'' {\em
  Physics of plasmas}, vol.~15, no.~5, p.~055502, 2008.

\bibitem{sengele2009microfabrication}
S.~Sengele, H.~Jiang, J.~H. Booske, C.~L. Kory, D.~W. Van~der Weide, and R.~L.
  Ives, ``Microfabrication and characterization of a selectively metallized
  $\text{W-band}$ meander-line $\text{TWT}$ circuit,'' {\em IEEE transactions
  on electron devices}, vol.~56, no.~5, pp.~730--737, 2009.

\bibitem{booske2011vacuum}
J.~H. Booske, R.~J. Dobbs, C.~D. Joye, C.~L. Kory, G.~R. Neil, G.-S. Park,
  J.~Park, and R.~J. Temkin, ``Vacuum electronic high power terahertz
  sources,'' {\em IEEE Transactions on Terahertz Science and Technology},
  vol.~1, no.~1, pp.~54--75, 2011.

\bibitem{gong2011experimental}
H.~Gong, Y.~Gong, T.~Tang, J.~Xu, and W.-X. Wang, ``Experimental investigation
  of a high-power $\text{Ka-band}$ folded waveguide traveling-wave tube,'' {\em
  IEEE transactions on electron devices}, vol.~58, no.~7, pp.~2159--2163, 2011.

\bibitem{armstrong2018compact}
C.~M. Armstrong, R.~Kowalczyk, A.~Zubyk, K.~Berg, C.~Meadows, D.~Chan,
  T.~Schoemehl, R.~Duggal, N.~Hinch, R.~B. True, {\em et~al.}, ``A compact
  extremely high frequency $\text{MPM}$ power amplifier,'' {\em IEEE
  Transactions on Electron Devices}, vol.~65, no.~6, pp.~2183--2188, 2018.

\bibitem{benford2007high}
J.~Benford, J.~A. Swegle, and E.~Schamiloglu, {\em High power microwaves}.
\newblock CRC press, Boca Raton, FL,USA, 2007.

\bibitem{gilmour1994principles}
A.~Gilmour, {\em Principles of traveling wave tubes}.
\newblock Artech House, Norwood, MA, USA, 1994.

\bibitem{pierce1947theory}
J.~Pierce, ``Theory of the beam-type traveling-wave tube,'' {\em Proceedings of
  the IRE}, vol.~35, no.~2, pp.~111--123, 1947.

\bibitem{pierce1951waves}
J.~Pierce, ``Waves in electron streams and circuits,'' {\em Bell System
  Technical Journal}, vol.~30, no.~3, pp.~626--651, 1951.

\bibitem{vainshtein1956electron}
L.~Vainshtein, ``Electron waves in retardation (slow-wave) systems. 1. general
  theory,'' {\em SOVIET PHYSICS-TECHNICAL PHYSICS}, vol.~1, no.~1,
  pp.~119--134, 1956.

\bibitem{sturrock1958kinematics}
P.~A. Sturrock, ``Kinematics of growing waves,'' {\em Physical Review},
  vol.~112, no.~5, p.~1488, 1958.

\bibitem{tamma2014extension}
V.~A. Tamma and F.~Capolino, ``Extension of the pierce model to multiple
  transmission lines interacting with an electron beam,'' {\em IEEE
  Transactions on Plasma Science}, vol.~42, no.~4, pp.~899--910, 2014.

\bibitem{othman2016theory}
M.~A. Othman, V.~A. Tamma, and F.~Capolino, ``Theory and new amplification
  regime in periodic multimodal slow wave structures with degeneracy
  interacting with an electron beam,'' {\em IEEE Transactions on Plasma
  Science}, vol.~44, no.~4, pp.~594--611, 2016.

\bibitem{jassem2020theory}
A.~Jassem, Y.~Lau, D.~P. Chernin, and P.~Y. Wong, ``Theory of traveling-wave
  tube including space charge effects on the circuit mode and distributed cold
  tube loss,'' {\em IEEE Transactions on Plasma Science}, vol.~48, no.~3,
  pp.~665--668, 2020.

\bibitem{dawson1983particle}
J.~M. Dawson, ``Particle simulation of plasmas,'' {\em Reviews of modern
  physics}, vol.~55, no.~2, p.~403, 1983.

\bibitem{tskhakaya2007particle}
D.~Tskhakaya, K.~Matyash, R.~Schneider, and F.~Taccogna, ``The particle-in-cell
  method,'' {\em Contributions to Plasma Physics}, vol.~47, no.~8-9,
  pp.~563--594, 2007.

\bibitem{antonsen1997christine}
T.~M. Antonsen~Jr and B.~Levush, ``Christine: A multifrequency parametric
  simulation code for traveling wave tube amplifiers.,'' tech. rep., Naval
  Research Lab, Washington, DC, USA, 1997.

\bibitem{antonsen1998traveling}
T.~Antonsen and B.~Levush, ``Traveling-wave tube devices with nonlinear
  dielectric elements,'' {\em IEEE transactions on plasma science}, vol.~26,
  no.~3, pp.~774--786, 1998.

\bibitem{vlasov2002simulation}
A.~N. Vlasov, T.~M. Antonsen, D.~P. Chernin, B.~Levush, and E.~L. Wright,
  ``Simulation of microwave devices with external cavities using
  $\text{MAGY}$,'' {\em IEEE transactions on plasma science}, vol.~30, no.~3,
  pp.~1277--1291, 2002.

\bibitem{wohlbier2002multifrequency}
J.~G. Wohlbier, J.~H. Booske, and I.~Dobson, ``The multifrequency spectral
  eulerian ($\text{MUSE}$) model of a traveling wave tube,'' {\em IEEE
  transactions on plasma science}, vol.~30, no.~3, pp.~1063--1075, 2002.

\bibitem{solntsev2015beam}
V.~A. Solntsev, ``Beam--wave interaction in the passbands and stopbands of
  periodic slow-wave systems,'' {\em IEEE Transactions on Plasma Science},
  vol.~43, no.~7, pp.~2114--2122, 2015.

\bibitem{chernyavskiy2016large}
I.~A. Chernyavskiy, T.~M. Antonsen, A.~N. Vlasov, D.~Chernin, K.~T. Nguyen, and
  B.~Levush, ``Large-signal 2-($\text{D}$ modeling of folded-waveguide
  traveling wave tubes,'' {\em IEEE Transactions on Electron Devices}, vol.~63,
  no.~6, pp.~2531--2537, 2016.

\bibitem{chernyavskiy2017modeling}
I.~A. Chernyavskiy, T.~M. Antonsen, J.~C. Rodgers, A.~N. Vlasov, D.~Chernin,
  and B.~Levush, ``Modeling vacuum electronic devices using generalized
  impedance matrices,'' {\em IEEE Transactions on Electron Devices}, vol.~64,
  no.~2, pp.~536--542, 2017.

\bibitem{jabotinski2019calculation}
V.~Jabotinski, D.~Chernin, T.~M. Antonsen, A.~N. Vlasov, and I.~A.
  Chernyavskiy, ``Calculation and application of impedance matrices for vacuum
  electronic devices,'' {\em IEEE Transactions on Electron Devices}, vol.~66,
  no.~5, pp.~2409--2414, 2019.

\bibitem{minenna2019recent}
D.~F. Minenna, A.~G. Terentyuk, F.~Andr{\'e}, Y.~Elskens, and N.~M. Ryskin,
  ``Recent discrete model for small-signal analysis of traveling-wave tubes,''
  {\em Physica Scripta}, vol.~94, no.~5, p.~055601, 2019.

\bibitem{marcuvitz1951waveguide}
N.~Marcuvitz, {\em Waveguide handbook}.
\newblock New York: McGraw-Hill, 1951.

\bibitem{felsen1994radiation}
L.~B. Felsen and N.~Marcuvitz, {\em Radiation and scattering of waves}.
\newblock John Wiley \& Sons, Hoboken, NJ, USA, 1994.

\bibitem{collin1990field}
R.~E. Collin, {\em Field theory of guided waves}.
\newblock John Wiley \& Sons, Hoboken, NJ, USA, 1990.

\bibitem{tsimring2006electron}
S.~E. Tsimring, {\em Electron beams and microwave vacuum electronics}.
\newblock John Wiley \& Sons, Hoboken, NJ, USA, 2007.

\bibitem{gilmour2011klystrons}
A.~Gilmour, {\em Klystrons, traveling wave tubes, magnetrons, crossed-field
  amplifiers, and gyrotrons}.
\newblock Artech House, Norwood, MA, USA, 2011.

\bibitem{pierce1947theoryTWT}
J.~Pierce, ``Theory of the beam-type traveling-wave tube,'' {\em Proceedings of
  the IRE}, vol.~35, no.~2, pp.~111--123, 1947.

\bibitem{forsythe1977computer}
G.~E. Forsythe, ``Computer methods for mathematical computations.,'' {\em
  Prentice-Hall series in automatic computation}, Englewood Cliffs, NJ, USA,
  1977.

\bibitem{williams1990overdetermined}
G.~Williams, ``Overdetermined systems of linear equations,'' {\em The American
  Mathematical Monthly}, vol.~97, no.~6, pp.~511--513, 1990.

\bibitem{anton2013elementary}
H.~Anton and C.~Rorres, {\em Elementary linear algebra: applications version}.
\newblock John Wiley \& Sons, Hoboken, NJ, USA, 2013.

\bibitem{mealy2019backward}
T.~{Mealy}, A.~F. {Abdelshafy}, and F.~{Capolino}, ``Backward-wave oscillator
  with distributed power extraction based on exceptional point of degeneracy
  and gain and radiation-loss balance,'' in {\em 2019 International Vacuum
  Electronics Conference (IVEC)}, pp.~1--2, Busan, South Korea, 2019, doi:
  10.1109/IVEC.2019.8745292.

\bibitem{mealy2019exceptional}
T.~Mealy, A.~F. Abdelshafy, and F.~Capolino, ``Exceptional point of degeneracy
  in a backward-wave oscillator with distributed power extraction,'' {\em
  Physical Review Applied}, vol.~14, no.~1, p.~014078, 2020.

\bibitem{mealy2020exceptional}
T.~Mealy, A.~F. Abdelshafy, and F.~Capolino, ``Exceptional point of degeneracy
  in linear-beam tubes for high power backward-wave oscillators,'' {\em
  arXiv:2005.08912}, 2020.

\bibitem{abdelshafy2018electron}
A.~F. Abdelshafy, M.~A. Othman, F.~Yazdi, M.~Veysi, A.~Figotin, and
  F.~Capolino, ``Electron-beam-driven devices with synchronous multiple
  degenerate eigenmodes,'' {\em IEEE Transactions on Plasma Science}, vol.~46,
  no.~8, pp.~3126--3138, 2018.

\bibitem{srivastava2000design}
V.~Srivastava, R.~G. Carter, B.~Ravinder, A.~Sinha, and S.~Joshi, ``Design of
  helix slow-wave structures for high efficiency $\text{TWT}$,'' {\em IEEE
  Transactions on Electron Devices}, vol.~47, no.~12, pp.~2438--2443, 2000.

\bibitem{pozar}
D.~M. Pozar, {\em Microwave engineering}.
\newblock John Wiley \& Sons, Hoboken, NJ, USA, 2011.

\bibitem{kosmahl1984twt}
H.~Kosmahl and J.~Peterson, ``A $\text{TWT}$ amplifier with a linear power
  transfer characteristic and improved efficiency,'' in {\em Proceedings of
  NASA Technology Memorandum 10th Communications Satellite Systems Conference},
  p.~762, Orlando, FL, USA, 1984.

\bibitem{feng2011study}
J.~Feng, D.~Ren, H.~Li, Y.~Tang, and J.~Xing, ``Study of high frequency folded
  waveguide $\text{BWO}$ with $\text{MEMS}$ technology,'' {\em Terahertz
  Science and Technology}, vol.~4, no.~4, pp.~164--180, 2011.

\bibitem{srivastava2018design}
A.~Srivastava and V.~L. Christie, ``Design of a high gain and high efficiency
  $\text{W}$-band folded waveguide $\text{TWT}$ using phase-velocity taper,''
  {\em Journal of Electromagnetic Waves and Applications}, vol.~32, no.~10,
  pp.~1316--1327, 2018.

\end{thebibliography}

\end{document}